\documentclass{article}

\usepackage{latexsym, amsmath, bbm}
\usepackage{amssymb}
\usepackage{amsfonts}
\usepackage{amstext}
\usepackage{multicol}
\usepackage{tabularx}
\usepackage{hyperref}
\usepackage{url}
\usepackage{appendix}
\usepackage{graphicx,stackengine,xcolor}
\usepackage{array,multirow}
\usepackage{footnote}
\usepackage{subcaption}
\usepackage{array}
\usepackage{censor}

\makesavenoteenv{table}

\newtheorem{Definition}{Definition}

\newcolumntype{L}[1]{>{\raggedright\arraybackslash}p{#1}}
\newcolumntype{C}[1]{>{\centering\arraybackslash}p{#1}}
\newcolumntype{R}[1]{>{\raggedleft\arraybackslash}p{#1}}
\newcolumntype{M}[1]{>{\centering\arraybackslash}m{#1}}

\newcommand{\D}{{\mathcal{D}}}
\newcommand{\GEO}{{\text{GEO}}}

\newcommand\Circle[1]{%
  \def\useanchorwidth{T}%
  \def\stacktype{L}%
  \stackon[0pt]{#1}{\scalebox{4.4}[1.7]{\textcolor{red}{$\bigcirc$}}}%
}

\title{The Geography and Election Outcome (GEO) Metric: an Introduction}

\begin{document}
\title{The Geography and Election Outcome (GEO) Metric:  an Introduction}

\author{Marion Campisi \\
\small Department of Mathematics \\[-.8ex]
\small San Jose State University \\[-0.8ex] 
\small \tt marion.campisi@sjsu.edu
\and 
Thomas Ratliff \\
\small Department of Mathematics  \\[-.8ex]
\small Wheaton College  \\[-.8ex]
\small \tt ratliff\_thomas@wheatoncollege.edu
\and
Stephanie Somersille \\
\small Somersille Math Education Services \\ [-.8ex]
\small \tt ssomersille@gmail.com
\and
Ellen Veomett\thanks{Following the convention in mathematics, author order is alphabetical.} \\
\small Dept of Mathematics and Computer Science \\[-0.8ex]
\small Saint Mary's College of California \\[-0.8ex]
\small \tt erv2@stmarys-ca.edu
\and
Short title: The GEO Metric: an Introduction
\and
}

\maketitle
%

\StopCensoring

\begin{abstract}
We introduce the Geography and Election Outcome (GEO) metric, a new method for identifying potential partisan gerrymanders.  In contrast with currently popular methods, the GEO metric uses both geographic information about a districting plan as well as district-level partisan data, rather than just one or the other. We motivate and define the GEO metric, which gives a count (a non-negative integer) to each political party.  The count indicates the number of previously lost districts which that party potentially could have had a 50\% chance of winning, without risking any currently won districts, by making reasonable changes to the input map.    We then analyze GEO metric scores for each party in several recent elections.  We show that this relatively easy to understand and compute metric can encapsulate the results from more elaborate analyses. 
\end{abstract}

\section{Introduction}

Partisan gerrymandering is an issue which has been adjudicated many times in recent years, including at the Supreme Court \cite{RuchoCommonCause}.  In these cases, the metrics used to identify partisan gerrymandering fall broadly into two categories.  The first category contains those that use data about a map to identify irregularly shaped districts and flag them as potential gerrymanders.   Possibly the most widely used of these map metrics is the Polsby Popper Ratio, which calculates a multiple of the ratio of the district's area to the square of its perimeter.  Thus, it effectively measures the irregularity of a district's boundary.  Other common map metrics are the Reock ratio (the ratio of a district's area to the area of the smallest disk containing the district), the Convex Hull ratio (the ratio of the area of the district to the area of its convex hull), and the Perimeter test (which simply sums the perimeters in all the districts)\cite{LWV_Intro_Metrics}.   But modern technology allows partisan demographers the possibility of creating hundreds of thousands of maps, all having reasonably shaped districts, and then selecting the most partisan among those.  Thus, looking for irregularly shaped districts is no longer an effective way of finding partisan bias in a map.  Technology also makes computation of boundaries ill-defined, depending on the level of map precision, as was discussed in \cite{DuchinTenner}.  This is not to mention the choice of type of sphere-to-plane map projection \cite{GerrymanderingJumble}, or the many other decisions made in computing shape metrics that greatly impact the analysis \cite{PA_Gerrymandering_Compactness}.  These issues have led to the introduction of metrics relying on election data instead.

Thus, the second typical category of metrics is those that use election outcome data.    Very generally, these metrics attempt to measure the ``packing and cracking'' that is widely understood to be how gerrymandering occurs.  ``Packing and cracking'' is present when a mapmaker ``packs'' her opponents into a small number of districts which are won with an overwhelming majority, and then ``cracks'' the remaining opponents among many districts in which they cannot gain a majority.  Perhaps the most common examples of metrics using election outcome data only are the Mean Median Difference and the Efficiency Gap.   The Mean-Median Difference calculates the median vote share among all districts, and subtracts from that median the average (the mean) of the vote shares among all districts.  The Efficiency Gap is based on the concept of a ``wasted vote;'' a vote is considered ``wasted'' if it was for a losing candidate or if it was a vote beyond the majority needed to win in a district.  The Efficiency Gap calculates the difference between two parties' wasted votes, and then divides by the total votes.  Other metrics using only election data include the Partisan Bias and the Declination; see \cite{LWV_Intro_Metrics} for descriptions of all of these ``election data'' metrics.  All of these metrics use nothing about the map, outside of how many votes each candidate received in each district.  They are not influenced at all by the locations of the voters, or the locations of the districts.

 In what follows we  define a new method, the Geography and Election Outcome (GEO) metric, which uses \emph{both} map and election outcome data to identify partisan gerrymanders.  Extensive analysis of maps drawn based on the 2020 census using the GEO metric is available at \censor{\url{https://www.the-geometric.com/}}.    We now provide an example which motivates the need to incorporate both the geographic information and the election outcome information in order to more accurately detect the presence of gerrymandering.
 
 \subsection{A motivating example}\label{subsec:MotivatingExample}
 
 Consider two states, State X and State Y, each with ten districts, two political parties, Party P and Party Q, and the election outcome data in Table \ref{StateXY}, where $V_{P}$ and $V_{Q}$ represent the vote shares for parties $P$ and $Q$, respectively.

\begin{table}[h]
\centering
\begin{tabular}{|c||C{1.3cm}|C{1.3cm}||C{1.3cm}|C{1.3cm}||}\hline
& \multicolumn{2}{|c||}{State $X$} & \multicolumn{2}{c||}{State $Y$} \\ \hline
District & $V_{P}$ & $V_{Q} $ &$V_{P}$ & $V_{Q}$   \\ \hline
1 & 10 \% & 90 \% &   62 \% & 38 \%   \\ \hline
2 & 10 \% & 90 \% & 62 \% & 38 \%    \\ \hline
3 & 46 \% & 54 \%  &  61 \% & 39 \%   \\ \hline
4 & 46 \% & 54 \%  &  61 \% & 39 \%   \\ \hline
5 & 61 \% & 39 \%  &10 \% & 90 \%     \\ \hline
6 &  61 \% & 39 \% &  10 \% & 90 \%   \\ \hline
7 & 46 \% & 54 \%  &  46 \% & 54 \%    \\ \hline
8 &   46 \% & 54 \% & 46 \% & 54 \%   \\ \hline
9 &   62 \% & 38 \%  &  46 \% & 54 \%   \\ \hline
10 &    62 \% & 38 \%  &  46 \% & 54 \%   \\ \hline
\end{tabular}
\caption{Two states with the same election outcomes.  $EG = 0$ for both states.}
\label{StateXY}
\end{table}
 
Aside from the district numbers, these have the exact same election outcome data and therefore will have the same results from a metric using only election data, such as the Efficiency  Gap.  Indeed, if we assume equal turnout in all districts, then the Efficiency Gap of both of these elections is 0.
 
\begin{figure}[h]
	\centering
	\subcaptionbox{State $X$ \label{StateX}}
         {\includegraphics[width=1 in]{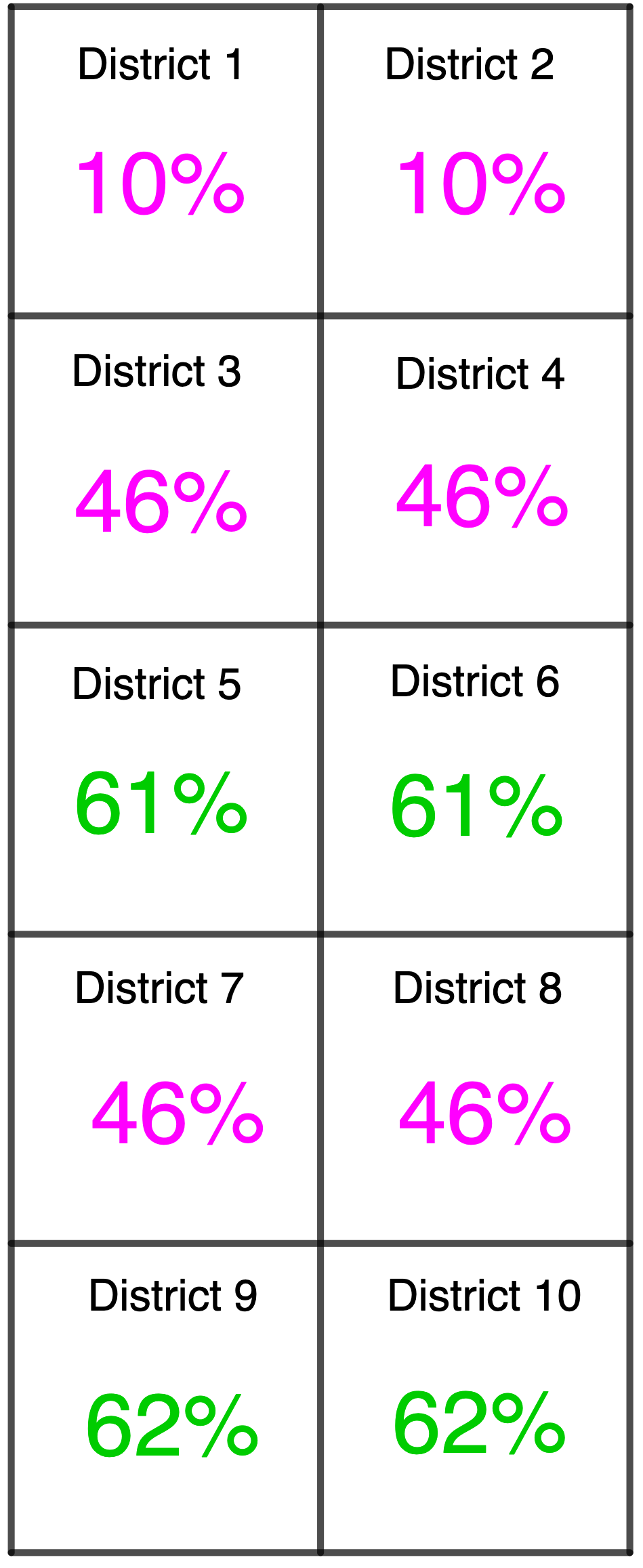}}
       \subcaptionbox{State $Y$ \label{StateY}}
         {\includegraphics[width=1 in]{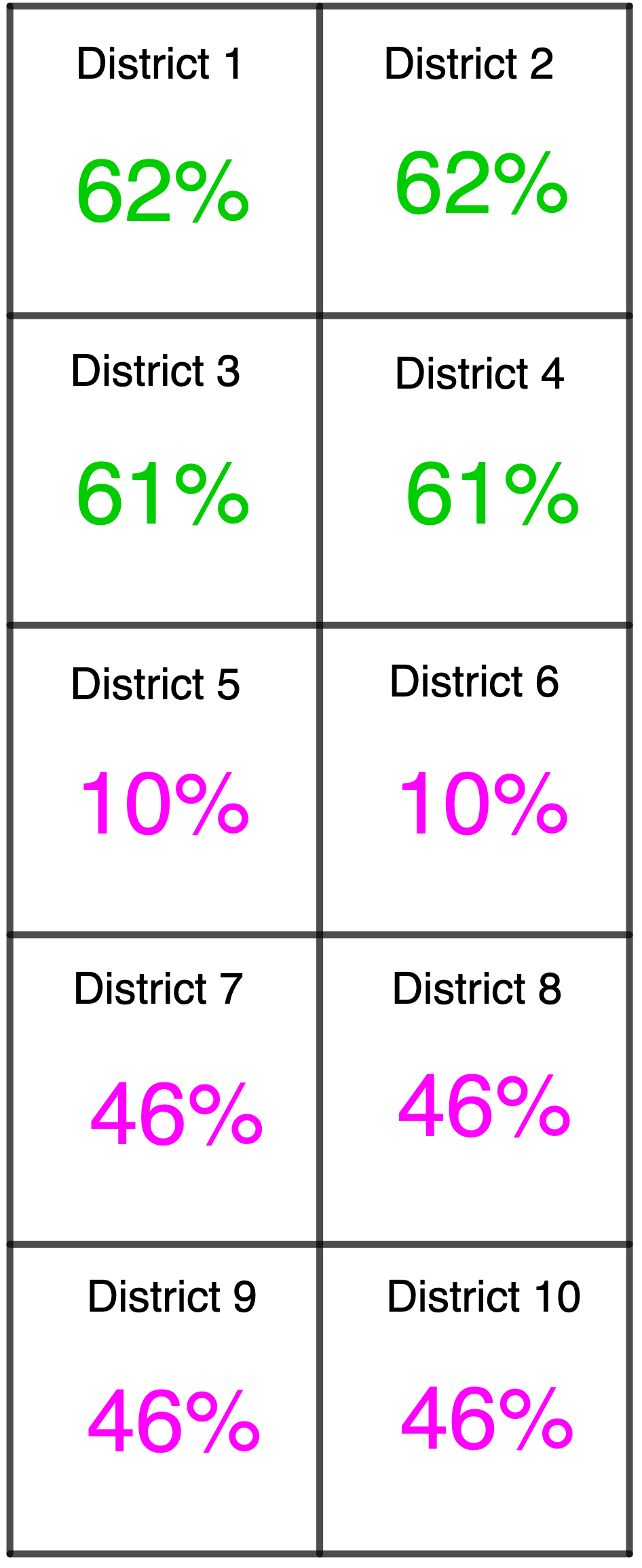}}
        \caption{Vote shares are for party $P$.}
        \label{BothStates}
\end{figure}

Now consider the maps in Figure \ref{BothStates} which correspond to State $X$ and State $Y$.  

We see that in State $X$, districts 3, 4, 7, and 8 appear to be potentially cracked for party $P$, as they are losses for $P$, have a vote share close to 50\%, and are adjacent to districts which are safe wins for party $P$.  That is, party $P$ has the possibility of improving its election outcome, based on the locations of the districts within the state.  On the other hand, in State $Y$, while districts 7, 8, 9, and 10 have the same vote shares for party $P$ as districts 3, 4, 7, and 8 in State $X$, their loss for party $P$ seems more an artifact of the lack of party $P$ voters in the southern part of the state than an intentional cracking.  Through this example, we can see that the \emph{location of the voters matters} when it comes to the potential presence  of packing and cracking.

In other words, partisan gerrymandering occurs when district lines are drawn so as to include or exclude voters in particular regions, resulting in a structural advantage for a particular political party.  This idea assumes that the lines could have been re-drawn so as to have a different outcome.  That is, certain districts have voters nearby that could have changed the outcome in that district.   In defining the GEO metric, we capture this missing aspect of election outcome data only methods: whether the ``packing and cracking'' detected via election outcome data is geographically realizable or is simply an artifact of the voter distribution within the state.  Indeed, in Section \ref{sec:TheAlgorithm}, we will see that for the Example in Figure \ref{BothStates}, the GEO metric score indicates a disadvantage for party $P$ in state $X$, but indicates no advantage for either party in state $Y$.

\subsection{An overview of the GEO metric}

The inputs for the GEO metric are both a districting plan $\D$ and  district-level partisan distribution $\Delta$. In this introductory paper, we assume there are just two parties; party $P$ and party $Q$.  In practice, the results from a statewide election are often used to determine the distribution $\Delta$. A score is given to each of the parties in the election, which we denote by 
\begin{equation*}
\GEO_P(\D,\Delta) \text{ or } \GEO_Q(\D,\Delta)
\end{equation*}
This score is in fact a count, as it corresponds to the number of districts a party lost that might have become competitive (for us, a 50/50 split, so that the party now has a 50\% chance of winning it), given small perturbations in the map,  without risking any currently held districts.  The GEO metric detects these new potential wins by considering vote share swaps with other districts with whom it shares a border.  Vote share swaps are limited so that a district's vote share does not fall into a probabilistically unlikely region, given the regional average vote share.   Along with the GEO score giving the count of newly competitive districts, we can list which districts became competitive through these vote share swaps, which districts won by party $X$ contributed to making another district newly competitive, and which districts lost by party $X$ contributed to making another district newly competitive. 

We note that the GEO metric is \emph{not} symmetric in the two parties.  That is, party $P$'s GEO  score is \emph{not} the negative of party $Q$'s GEO score.  We view this as a benefit, in that it recognizes that party $P$'s voters may distribute themselves throughout a state very differently from party $Q$'s voters.  We agree with DeFord et al. in their argument that ``there are serious obstructions to the practical implementation of
symmetry standards'' and that methods centered on varying districting lines (rather than votes) are better at assessing the presence of partisan map manipulation \cite{ImplementingPartisanSymmetry}.

It is worth noting that the GEO metric is not the \emph{only} metric which uses both geographic and partisan data in order to detect gerrymandering; the Partisan Dislocation \cite{PartisanDislocation} and the Gerrymandering Index \cite{DukeNC} are other such metrics.  However, the GEO metric is much easier to compute than the Partisan Dislocation, which requires extremely fine data on the location of voters within the state and their partisan leanings.  The GEO metric is also deterministic, unlike the Gerrymandering Index, which relies on the creation of an ensemble of districting maps.

  The paper is structured as follows:  Section \ref{section:defns} contains relevant definitions and background.  In Section \ref{section:algorithm} we describe the algorithm by which we compute the GEO metric for a given districting plan and election outcome data.  In Section \ref{section:real} we analyze maps from the 2011 redistricting cycle to illustrate that the GEO metric results align with what more in depth analyses of these maps indicated, in Section \ref{section:newdata} we analyze maps from the 2021 redistricting cycle, and in Section \ref{section:analysis} we give a mathematical description and discussion of the GEO metric. In Section \ref{section:MCMC} we explore the use of the GEO metric on ensembles of maps.  Finally, in Section \ref{section:CCT} we highlight some caveats, clarifications, and takeaways.  

\section{Definitions}
\label{section:defns}

\begin{figure}[h]
	\centering
	\subcaptionbox{Districting graph for state $X$ and party $P$. \label{GraphStateX}}
         {\includegraphics[width=1.2 in]{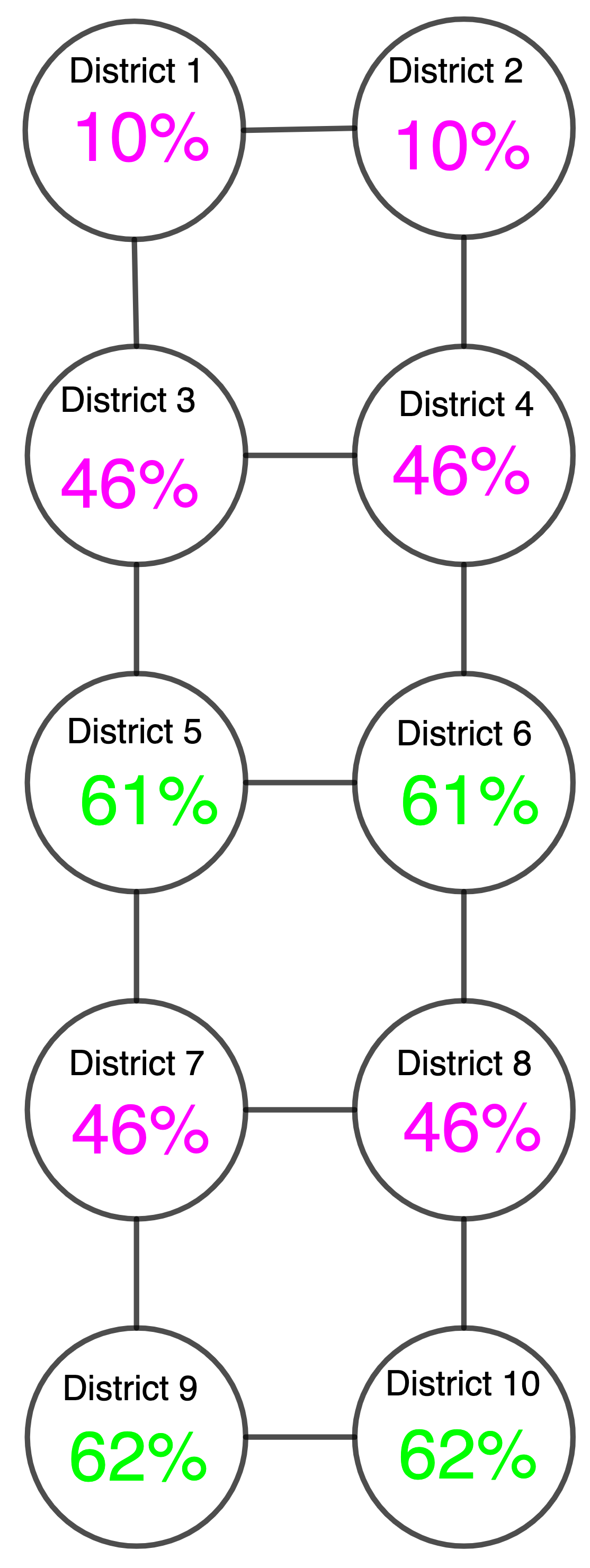}} \hspace{1 cm}
       \subcaptionbox{Districting graph for state $Y$ and party $P$. \label{GraphStateY}}
         {\includegraphics[width=1.2 in]{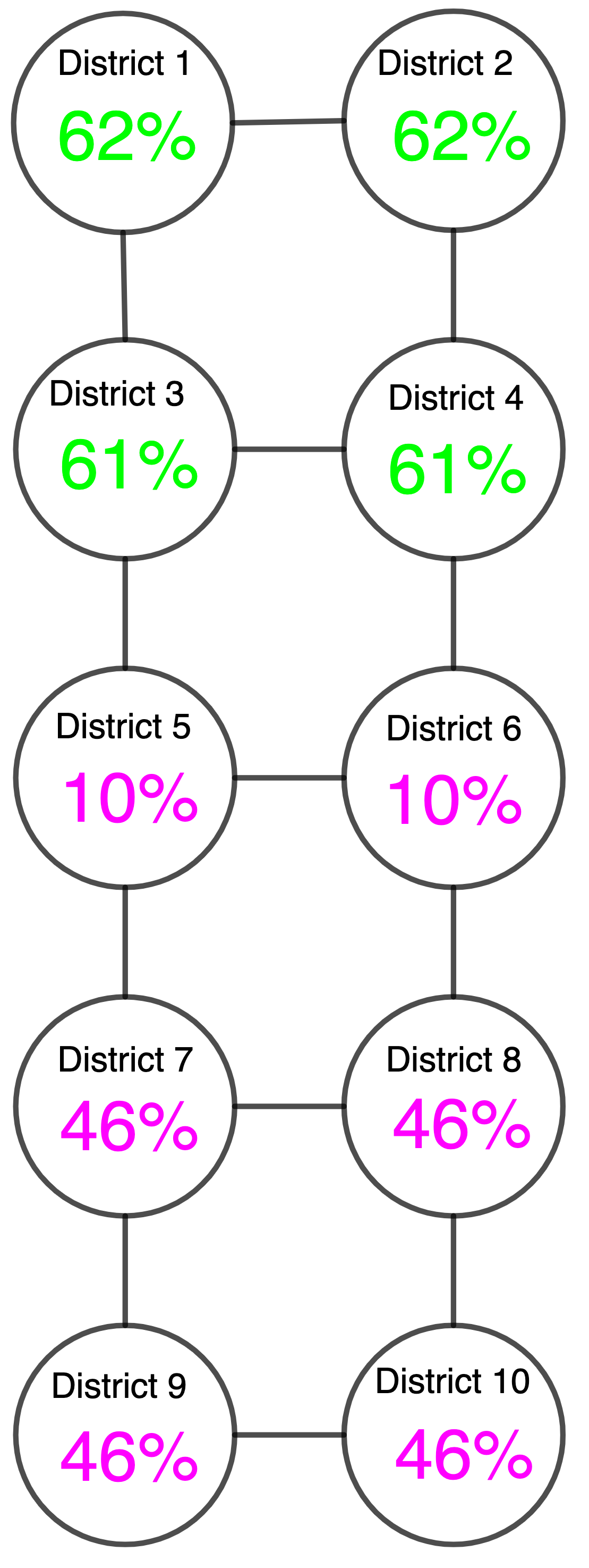}}
        \caption{}
        \label{BothGraphs}
\end{figure}

Here, we introduce the notation that will be used throughout.   We start with  a districting plan $\mathcal{D}$, consisting of districts $D_1, D_2, \dots, D_n$ and partisan distribution $\Delta$.   We say that districts $D_i$ and $D_j$ \emph{share a boundary} if $i \not=j$, and the intersection of $D_i$ and $D_j$ is a 1-dimensional shape of positive length.  This is sometimes referred to as \emph{rook adjacency}, as opposed to \emph{queen adjacency} which also considers districts to be neighbors if they share a single point.  The  \emph{districting graph} is the dual graph of the districting map.  That is, the vertices of our graph are $D_1, D_2, \dots, D_n$ and we say that $(D_i,D_j)$ is an edge if  districts $i$ and $j$ share a boundary.  A districting graph from the states in Figure \ref{BothStates} can be seen in Figure \ref{BothGraphs}

Each district $D_i$ has a vote share for party $P$, which we denote by $V_i$.   Each district is put into one of four categories, depending on $V_i$.

\begin{Definition}

\emph{Loss}: We say a district is a \emph{loss} for party $P$ if party $P$ wins some percentage of the vote share less than 50\%:  $V_i < 0.5$.

\emph{Unstable win}: A district is an \emph{unstable win}  for party $P$ if party $P$ wins some percentage of the vote share which is larger than 50\% but smaller than some fixed parameter $w$.  For the purposes of our calculations in this introductory paper, we will set $w =  0.55$.  Thus, a district $D_i$ is an unstable win if  $V_i \in (0.5, w)$, with $w = 0.55$ in the examples presented in this paper.

\emph{Stable win}: A district is a \emph{stable win} for party $P$ if party $P$ wins some percentage of the vote which is at least $w$.   For the calculations in this introductory paper, this implies  $V_i \geq 0.55$.

\emph{Even split}: If the vote share for party $P$ in the district is precisely 50\% we say that district is \emph{evenly split}: $V_i = 0.5$. Note, we do not expect districts to naturally achieve a precisely 50\% vote share. This designation will be used in what follows to calculate the GEO metric.
\end{Definition}

We let $N_i = \left\{j \not= i: D_j \text{ shares a boundary with } D_i \right\}$ denote the indices of $D_i$'s neighboring districts.  We calculate district $D_i$'s \emph{average neighborhood vote share} by averaging the vote shares of $D_i$, along with all of its neighbors in the districting graph:
\begin{equation*}
A_i = \frac{V_i + \sum_{j \in N_i} V_j}{1+\left| \{j\in N_i\} \right|} \quad \quad \quad \quad i = 1, 2, \dots, n
\end{equation*}
The value $A_i$ can be considered to be the regional support for party $P$ in the region surrounding district $D_i$.

We let $\sigma$ be the standard deviation of the set of all $A_i$ for each district in the map:

\begin{equation*}
\sigma = \sqrt{\frac{\sum_{i=1}^n \left(\mu- A_i\right)^2}{n}} \quad \quad \quad \quad \quad \text{ where } \quad \quad  \mu = \frac{\sum_{i=1}^n A_i}{n}
\end{equation*}
Since $\sigma$ is the standard deviation of the regional average vote shares $A_i$, we expect it to be smaller than the standard deviation of the $V_i$s.

\begin{Definition}

We define district $D_i$'s \emph{shareable vote share} $S_i$ to be the vote share that $D_i$ has available to swap with neighbors, according to the algorithm we define for the GEO metric. For stable wins and losses this figure is the vote share that can be swapped without changing the district's  classification, and without it dropping to the level of one standard deviation below its average neighbor vote share.

That is, for a losing district,
\begin{equation*}
S_i  = \text{max}\left\{0, V_i - (A_i - \sigma)\right\}
\end{equation*}

And for a winning district,
\begin{equation*}
S_i = \text{max}\left\{0, \text{min}\left\{V_i-w,V_i-( A_i- \sigma)\right\}\right\}
\end{equation*}

\end{Definition}

Districts $D_i$ which are unstable wins are not altered by the GEO algorithm and have no shareable vote shares; thus, for those districts, $S_i = 0$.

For consistency and ease of presentation, we follow the convention of other authors (as in \cite{Competitiveness}) and use $w = 0.55$ as the lower bound for a safely won district in our calculations.   See Section \ref{section:CCT} for a list of topics for further research including varying $w$ based on state specific considerations.

To recap, the `stable win’ districts are districts with vote shares $V$ in the range from $0.55 \leq V \leq 1$. The limit 0.55 is intended to keep the districts `safe wins' after the vote share swap.   Remember that the goal of the algorithm calculating the GEO metric is to see how a party's outcome can be improved; having districts which were previously sure wins become an unstable win is not an improvement, which is why we have the 0.55 limit.  The `unstable win’ districts are those with vote shares $V$ such that $0.5 < V < 0.55$. They are not altered by the GEO algorithm so they stay in this category. We define districts as ``evenly split" if the vote share for each party is 50$\%$. These are the newly competitive districts in which each party has a 50/50 chance of winning.  The algorithm calculating the GEO metric changes previously lost districts to evenly split, so that the GEO metric number for party $P$ is interpretable as the number of districts that, under reasonable changes to the map, party $P$ now has a 50/50 chance of winning.  For researchers who may be interested in adjusting that lower bound to be potentially less than 50$\%$, we encourage them to use the code at www.the-geometric.com, as we've made that an easily changeable parameter.  For the purposes of our calculations in this paper, we use ``evenly split" to mean that the vote share is precisely 50$\%$.

\section{The GEO Metric} \label{section:algorithm}

The algorithm which calculates the GEO metric swaps vote shares between neighboring districts in a manner that is beneficial for the party in consideration, which we shall call party $P$.    As stated in Section \ref{section:defns}, we consider two distinct districts to be neighbors if they share a boundary whose 1-dimensional length is positive.  This in turn implies that their corresponding vertices in the districting graph share an edge.   Vote shares are swapped between neighboring districts in order to turn a lost district into a  district which is evenly split.  Specifically, vote shares are swapped in order to increase the vote share in a lost district to exactly 50\%, our even split or competitive designation.  We only move vote shares out of a district which is either a loss or a stable win, as districts categorized as an ``unstable win''  are unlikely to represent an entrenched bias.  We do not allow so many vote shares to be moved out of a safely won district so as to make it anything but a safely won district after the movement.  That is, after swapping vote shares out of a safely won district, we require that the district keep a vote share of at least $w$.  We also do not allow so many vote shares to be moved out of a losing district \emph{or} a safely won district so as to make the vote share for party $P$ drop below the regional average, minus one standard deviation of regional averages.  That is, using the notation in Section \ref{section:defns}, we do not allow a district's vote share to drop below $A_i - \sigma$; a value which is statistically reasonably close to the district's current neighborhood vote share.  Finally, when vote shares are swapped into district $D_i$, they are swapped in from all its neighboring districts \emph{proportional to their shareable votes}.  That is, we let $C_i$ be the vote share that district $D_i$ needs to become evenly split, and we let $T_i$ be the vote shares that can be transferred in from $D_i$'s neighbors:
\begin{equation*}
T_i = \sum_{j \in N_i} S_j 
\end{equation*}
Then if we have $T_i \geq C_i$, neighboring district $D_j$ swaps 
\begin{equation*}
S_j \cdot \frac{C_i}{T_i}
\end{equation*}
vote shares for party $P$ into district $D_i$ (while district $D_i$ swaps \emph{out} $S_j \cdot \frac{C_i}{T_i}$ vote shares for party $Q$ into district $D_j$).

We then count the number of districts that party $P$  lost which are now evenly split.  That count will be an indication of how many \emph{more} districts party $P$ ``could have won'' with 50\% probability, in addition to all of the districts it already did win.    We emphasize that the purpose of this algorithm is \emph{not} to find the movement of vote shares that would maximize the GEO score for party $P$.    Rather, we would like to notice any places where it seems \emph{likely} that a revision of district lines could have benefitted party $P$.

It is worth noting that it is \emph{vote shares} that are swapped between districts, rather than \emph{number of votes}.  The reason for this is because, while districts are drawn to have the same population, they are not drawn to have the same citizen voting age population and also turnout between districts can vary wildly (see \cite{2018arXiv180105301V} for how turnout can vary, as well as an example of how uneven turnout can skew the calculation of the Efficiency Gap).  Because of this uneven turnout, a single voter in one district (with low turnout) can represent a much higher percentage of the population in their district than a voter in another district (with high turnout).  Thus, swapping \emph{voters} between districts would correspond to swapping unequal populations.  However, swapping \emph{vote shares} corresponds to swapping the same represented population.

It remains to describe the details of the algorithm that swaps vote shares from one district into a neighboring district.  The algorithm is based in the intuitive idea that, to find gerrymandering, we look for where we think it is most likely.  That is, we look for districts that party $P$ lost, but which are in a region in which party $P$ has the highest vote share.

\subsection{The Algorithm}\label{sec:TheAlgorithm}

We describe here the details of the algorithm which calculates the GEO metric.  For those interested in calculating the GEO metric, the authors have made the Python code available at \censor{\url{https://www.the-geometric.com/}. } 

 For each district $D_i$, let $A_i$ be the average vote share for party $P$ among that district and all of its neighbors.  Thus, if a district is in a region in which party $P$ is very popular, then this average should be high.   In general, the higher this average, the more we would expect party $P$ to win districts in the area.  Then re-order the districts\footnote{It is statistically extremely improbable that two districts would have the same neighborhood average vote share $A_i$ in real-world data.  But if this were to happen, our Python code implementing the GEO metric would put the district which appears earlier in the data set, earlier in the ordering of $D_1, D_2, \dots, D_n$.} $D_1, D_2, \dots$ so that
\begin{equation*}
A_1 \geq A_2 \geq A_3 \geq \cdots
\end{equation*}

With this ordering, we do the following:
\begin{enumerate}
\item  In order $i=1, 2, \dots, n$, consider district $D_i$
\item  If that district was won by party $P$, we don't need to do anything further.  Increase $i$ and go back to step (1).
\item  Otherwise, that district was lost by party $P$.  Let $C_i$ be the amount of vote shares that district $D_i$ needs in order to become evenly split:  $C_i = 0.5 - V_i$. Let $T_i$ be the vote shares that can be transferred in from $D_i$'s neighbors:
\begin{equation*}
T_i = \sum_{j \in N_i} S_j
\end{equation*}
If $T_i < C_i$, $D_i$'s neighbors don't have enough vote shares for party $P$ in order to make party $P$ evenly split in that district.  Increase $i$ and go back to step (1). 
\item Otherwise, $D_i$'s neighbors \emph{do} have enough vote shares for party $P$ in order to make party $P$ evenly split in that district: $T_i \geq C_i$.  For each neighbor $D_j$ of $D_i$, that neighbor swaps out $S_j \cdot \frac{C_i}{T_i}$ vote shares for party $P$, and swaps in  $S_j \cdot \frac{C_i}{T_i}$ vote shares from party $Q$ from district $D_i$.  
\begin{enumerate}
\item District $D_i$'s vote share is thus updated to be $V_i = 0.5$
\item District $D_i$'s neighbor's vote shares are updated to be $V_j -S_j \cdot \frac{C_i}{T_i}$, and their shareable vote shares (described in Section \ref{section:defns}) are updated similarly.
\end{enumerate}
\item Increase $i$ and go back to step (1).  
\end{enumerate}

\begin{figure}[h]
	\centering
	\subcaptionbox{Initial party $P$ vote shares.  $A_i$ is average neighbor vote share\label{fig:Initial}}
         {\includegraphics[width = 0.7 in]{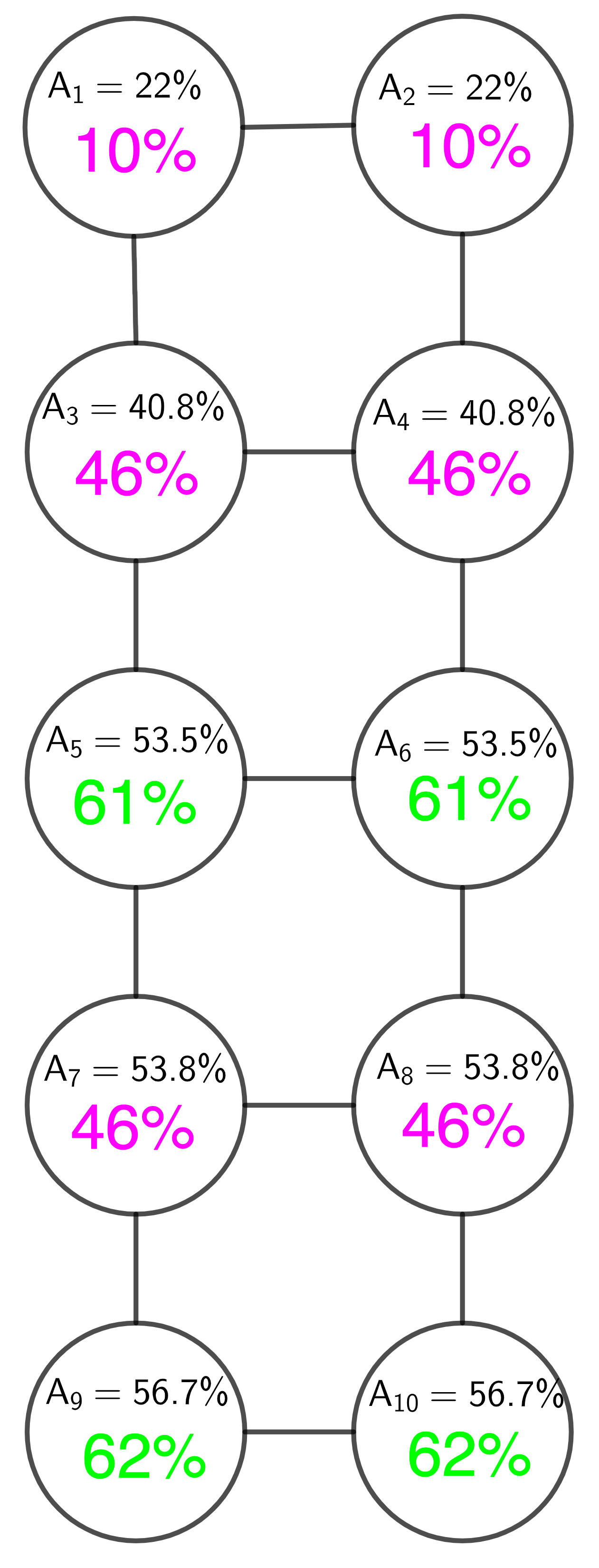}} \hspace{1 cm}
       \subcaptionbox{Direction of first vote share swap \label{fig:Motion1}}
         {\includegraphics[width=0.7 in]{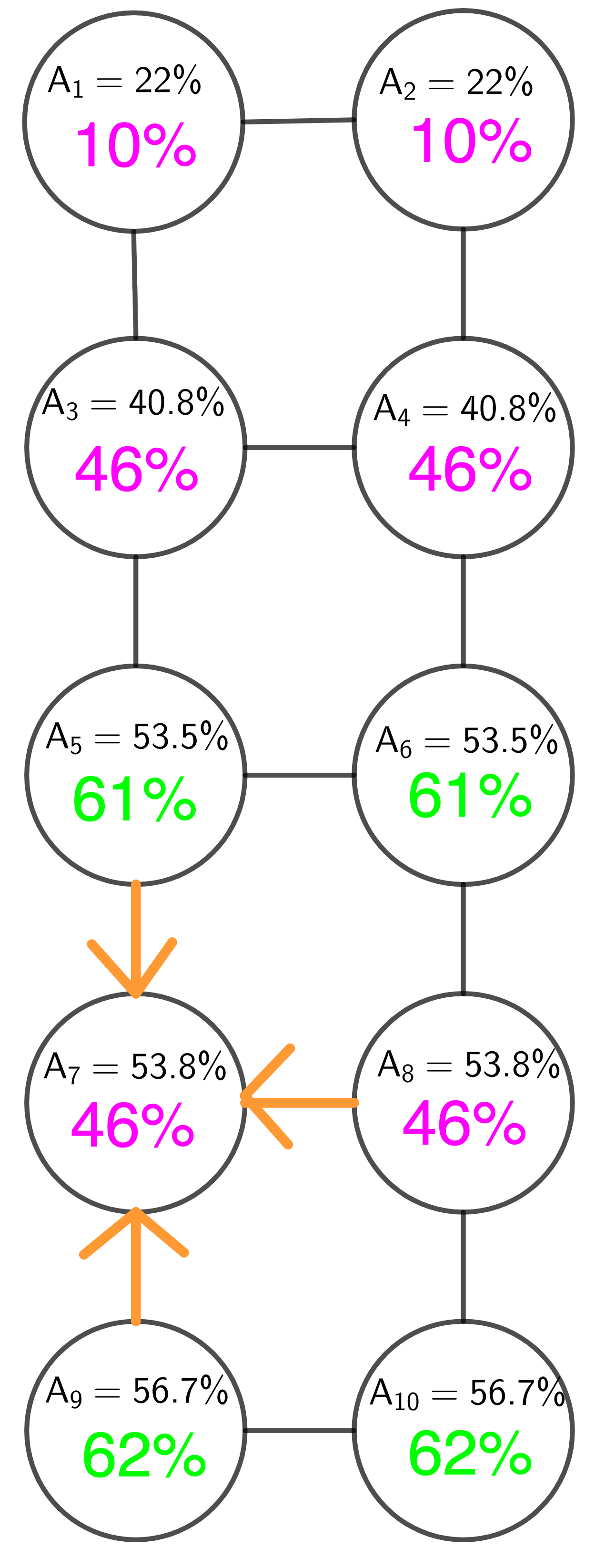}} \hspace{1 cm}
         \subcaptionbox{Outcome of first vote share swap \label{fig:Motion2}}         
         {\includegraphics[width=0.7 in]{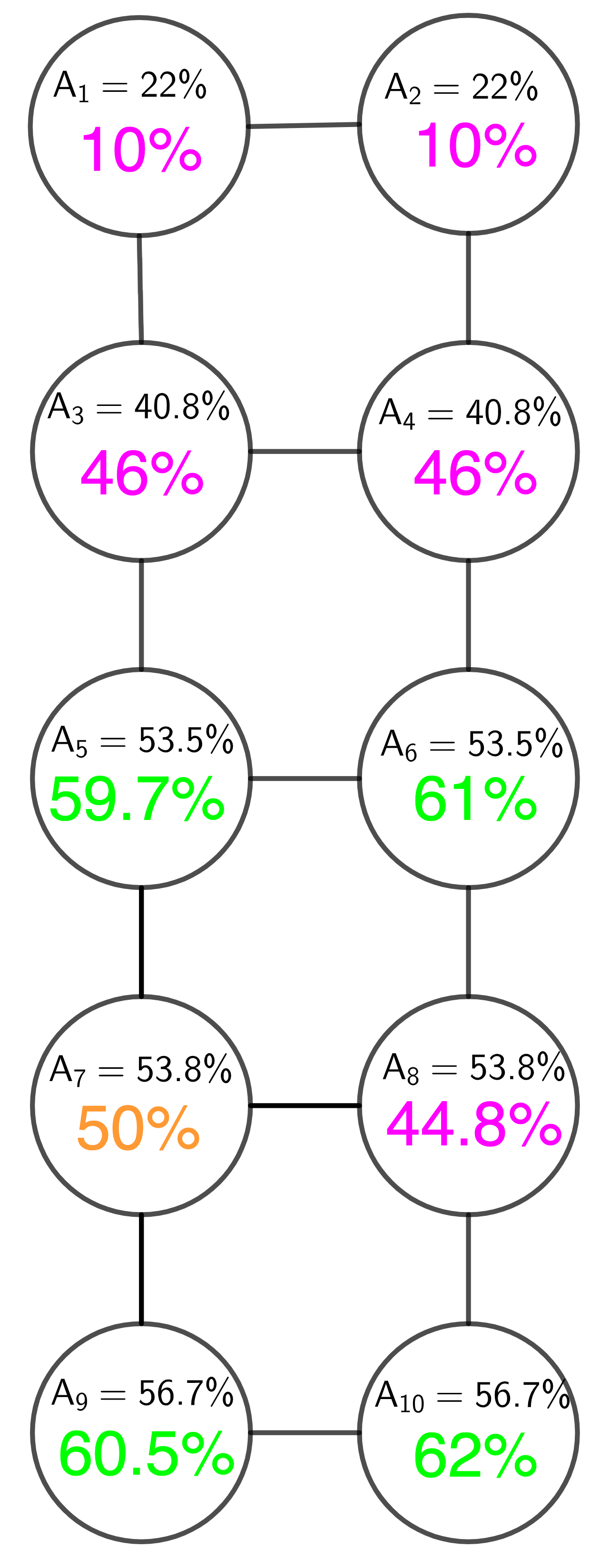}} \hspace{1 cm}

         \subcaptionbox{Direction of second vote share swap \label{fig:Motion3}}
         {\includegraphics[width=0.7 in]{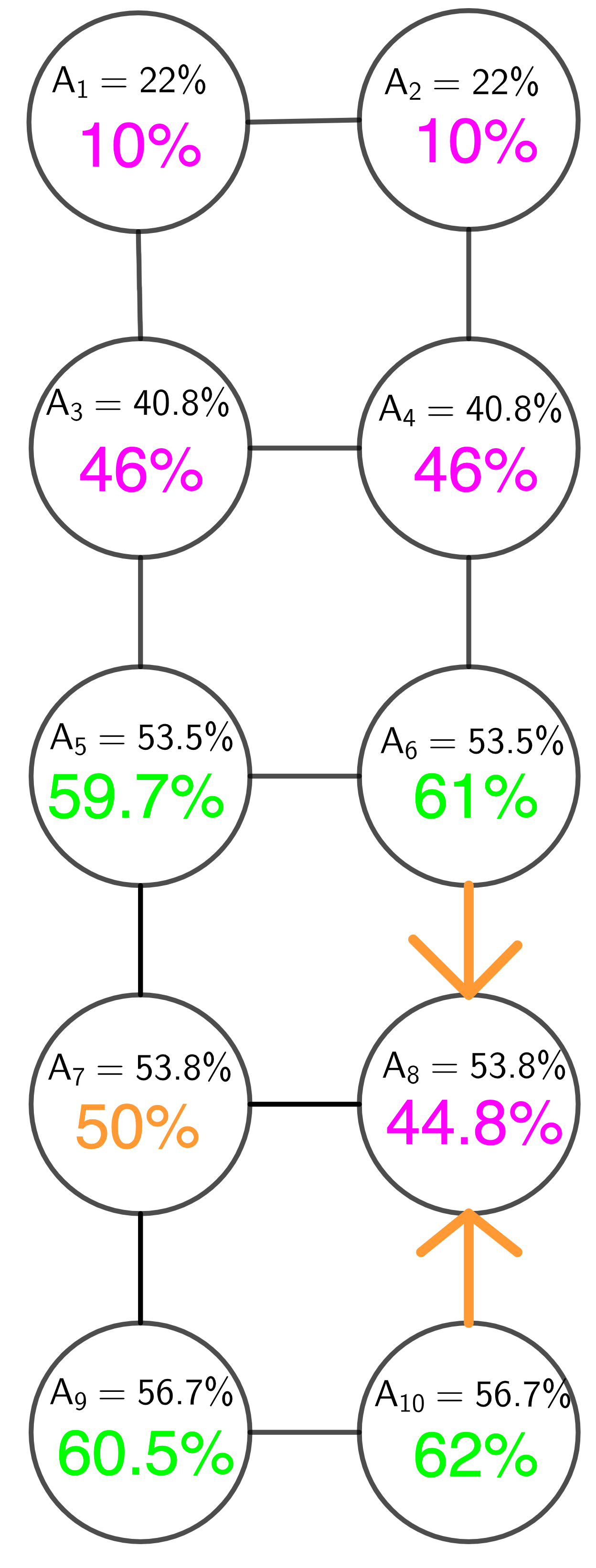}} \hspace{0.3 cm}
         \subcaptionbox{Outcome of second vote share swap \label{fig:Motion4}}
         {\includegraphics[width=0.7 in]{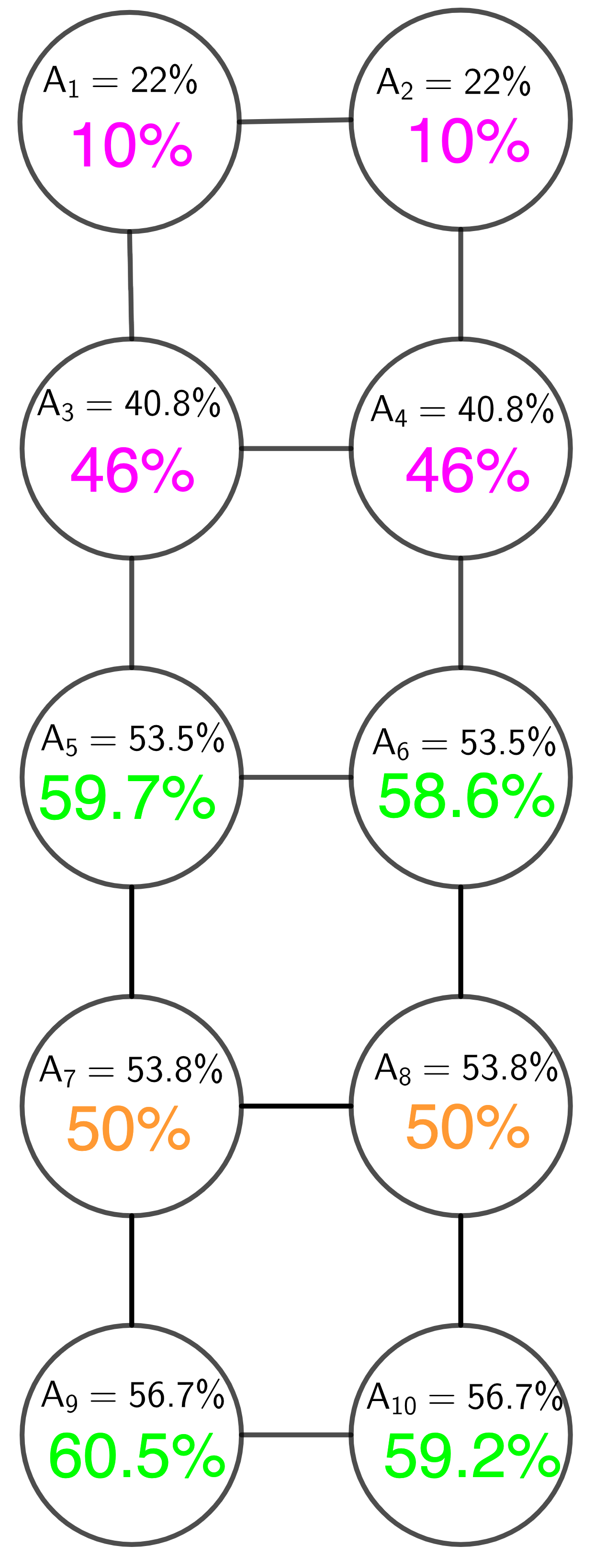}} \hspace{0.3 cm}
          \subcaptionbox{Direction of third vote share swap \label{fig:Motion4}}
         {\includegraphics[width=0.7 in]{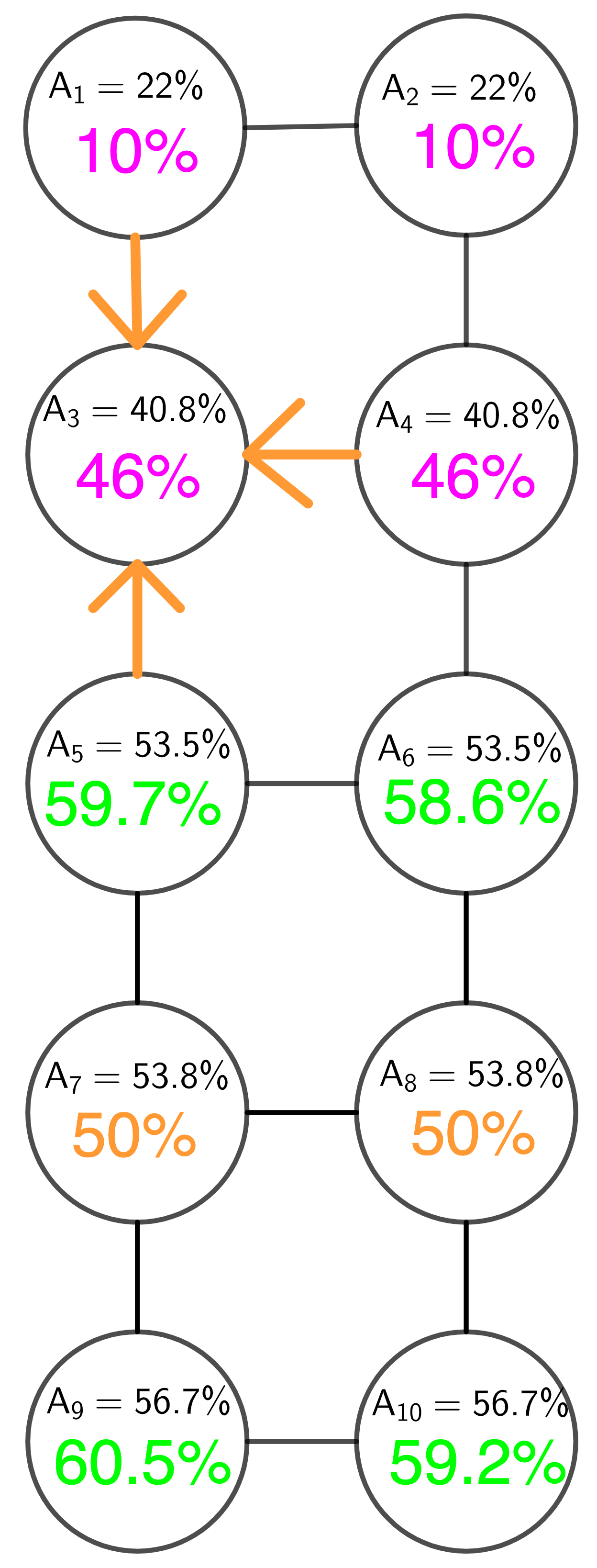}} \hspace{0.3 cm}
          \subcaptionbox{Outcome of third vote share swap \label{fig:Motion4}}
         {\includegraphics[width=0.7 in]{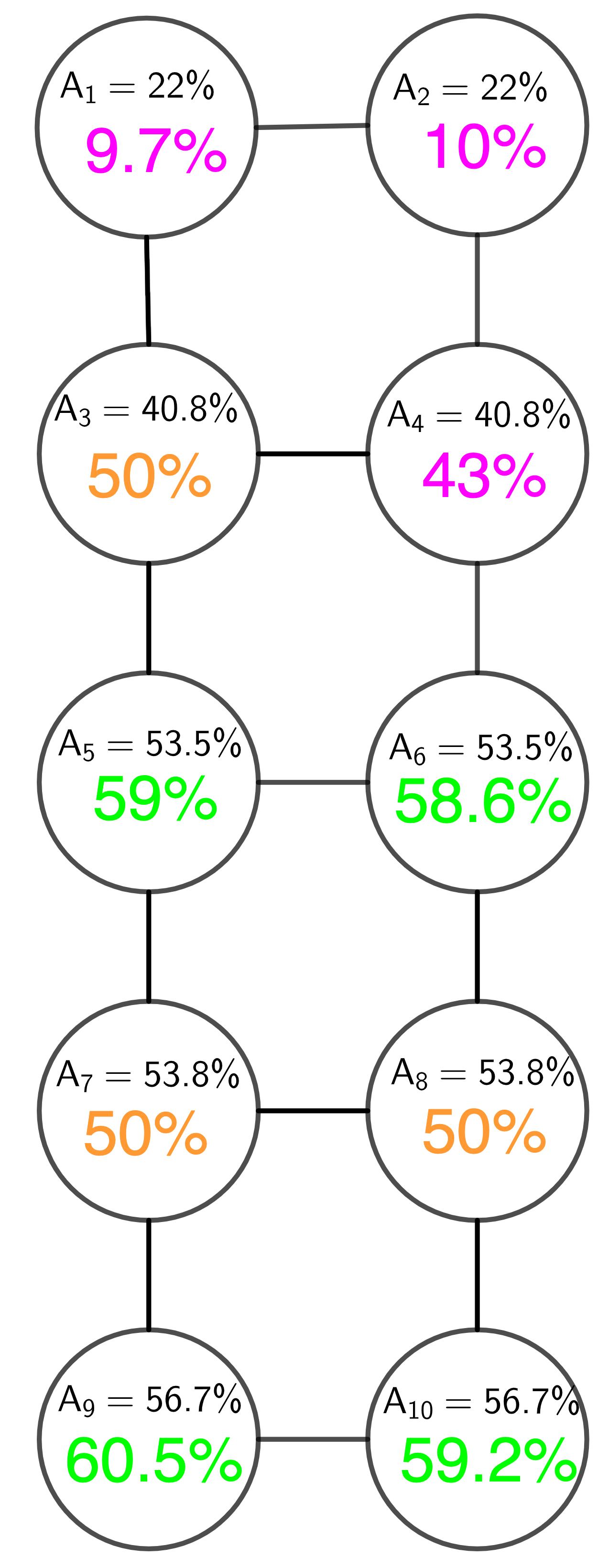}}
        \caption{Here, $\GEO_{P} = 3$.  (Party $P$ is the green party).}
        \label{fig:AlgorithmSteps}
\end{figure}

The value $\GEO_P$ for this map and election outcome is then the number of districts which are newly competitive\footnote{To avoid the awkward phrase ``newly evenly split'' we use the phrase ``newly competitive,'' which we more formally introduce in our district categories in the following paragraph.} after the algorithm has gone through each district.  As an example of the algorithm in action, we consider the sample state $X$ from Section \ref{subsec:MotivatingExample} whose districting graph appears in Section \ref{section:defns}.  The steps of the algorithm calculating the GEO metric can be seen visually in Figure \ref{fig:AlgorithmSteps}

At this point, we can also categorize some of the districts into: newly competitive, contributing stable wins, and contributing losses:
\begin{enumerate}
\item  If a district was previously a loss for party $P$ but was made evenly split by the algorithm calculating $\GEO_P$, we call that district  ``newly competitive.''
\item  If a district was won by party $P$ and had vote shares transferred out of it in order to make another district newly competitive during the algorithm, we call that district a ``contributing stable win.''
\item  If a district was lost by party $P$ and had vote shares transferred out of it in order to make another district newly competitive during the algorithm, we call that district a ``contributing loss'' district. 
\end{enumerate}

Recall that $A_i$ is the average vote share for party $P$ among district $D_i$ and
all of its neighbors. Thus in general, the larger $A_i$, the more we would expect
party $P$ to win district $D_i$. The ordering of districts by $A_i$ is not intended to maximize party $P$’s GEO score. The districts are ordered according to how much one would expect party $P$ to win each district.

In the example from Figure \ref{fig:AlgorithmSteps}, we show the steps of the algorithm calculating the GEO metric for party $P$  for state $X$ from Section \ref{subsec:MotivatingExample}.  In this example, we can see that Districts 7, 8, and 3 are newly competitive for party $P$, Districts 9 and 10 are contributing stable wins, and Districts 4 and 1 are contributing loss districts.  We won't show the steps for these calculations here, but we do note that party $Q$'s GEO score for state $X$ is 0. This pair of GEO scores captures the fact that the authors have drawn this map to improve party Q's outcome (since party $Q$ has very little room for improvement).  Whereas, for state $Y$ in Section  \ref{subsec:MotivatingExample}, party $P$'s GEO score is 3, and party $Q$'s GEO score is 2, indicating an absence of partisan gerrymandering, since both parties have essentially the same room for improvement on their current outcome.  We will come back to these examples in Section \ref{section:MCMC}.

We note here that, while the GEO metric counts the number of newly competitive districts, and thus indicates how many additional districts a party potentially could have won, the GEO metric score is \emph{not} intended to count the number of additional districts a party \emph{should} have won.  It is unreasonable that a party would win \emph{all} of its newly competitive districts.  Rather, it would be more reasonable to say that, because the newly competitive districts are evenly split, party $P$ could have won approximately $\GEO_P/2$ additional districts (beyond the districts they already won), with reasonable changes to the current map.  More importantly, the GEO score indicates the flexibility that a party has in improving its outcome.  If one party has a lot of flexibility to improve its outcome, while another has just a little or even none at all, this would indicate influence by the mapmaker to benefit the party which has little or no ability to improve its outcome.

\section{Maps from the 2011 Redistricting Cycle}
\label{section:real}

In this section, we show the results of the GEO metric analysis on the 2011 North Carolina Congressional districting map, the 2011 Pennsylvania congressional districting map and the Colorado's 2013-enacted map.   We’ve chosen these maps because North Carolina and Pennsylvania are largely understood to have been intentionally gerrymandered while it has recently been argued that Colorado does not have effective partisan manipulation \cite{ColoradoInContext}. We use the 2011 maps here to illustrate that the GEO captures the conclusions of thorough analyses. Indeed, the Pennsylvania State Supreme court declared that Pennsylvania's map violated the state constitution \cite{LeageWomenPA}.  And North Carolina's congressional redistricting map was struck down by the Supreme Court of the United States as an unconstitutional racial gerrymander \cite{CooperHarris}. 

For each state, we've chosen elections for statewide or national offices to determine the partisan distribution $\Delta$ because such elections are reasonable stand-ins for party preference.  As with all metrics using election data, those using the GEO Metric will want to take into consideration which elections or other partisan indicators they use to represent party preference.

For this introductory paper, we focus on the two-party calculation and analysis of elections with two parties.  Thus, all data in this section (Section \ref{section:real}) and in Section \ref{section:MCMC} have omitted votes that are not for the Democratic or Republican candidates.

Each table shows the GEO score for each party; the \emph{newly competitive districts},  and the \emph{contributing districts}, i.e. those that shared votes to contribute to at least one district becoming competitive. The ``newly competitive'' districts are ordered in the order they are analyzed: from largest to smallest average neighborhood vote share $A_i$.  The contributing districts are categorized as either contributing stable wins (districts whose initial vote shares were more than 55\%) or as contributing losses (districts whose initial vote shares were less than 50\%). In each of those contributing district categories, the districts are ordered by the total vote shares they swapped with other districts (from highest vote share swap to lowest). Again, we remind the reader that this algorithm is not intended to maximize the GEO score but to find a `reasonable’ number of districts that might have become competitive without perturbing any district too much.

\begin{figure}[h]
\centering
\includegraphics[width=3 in]{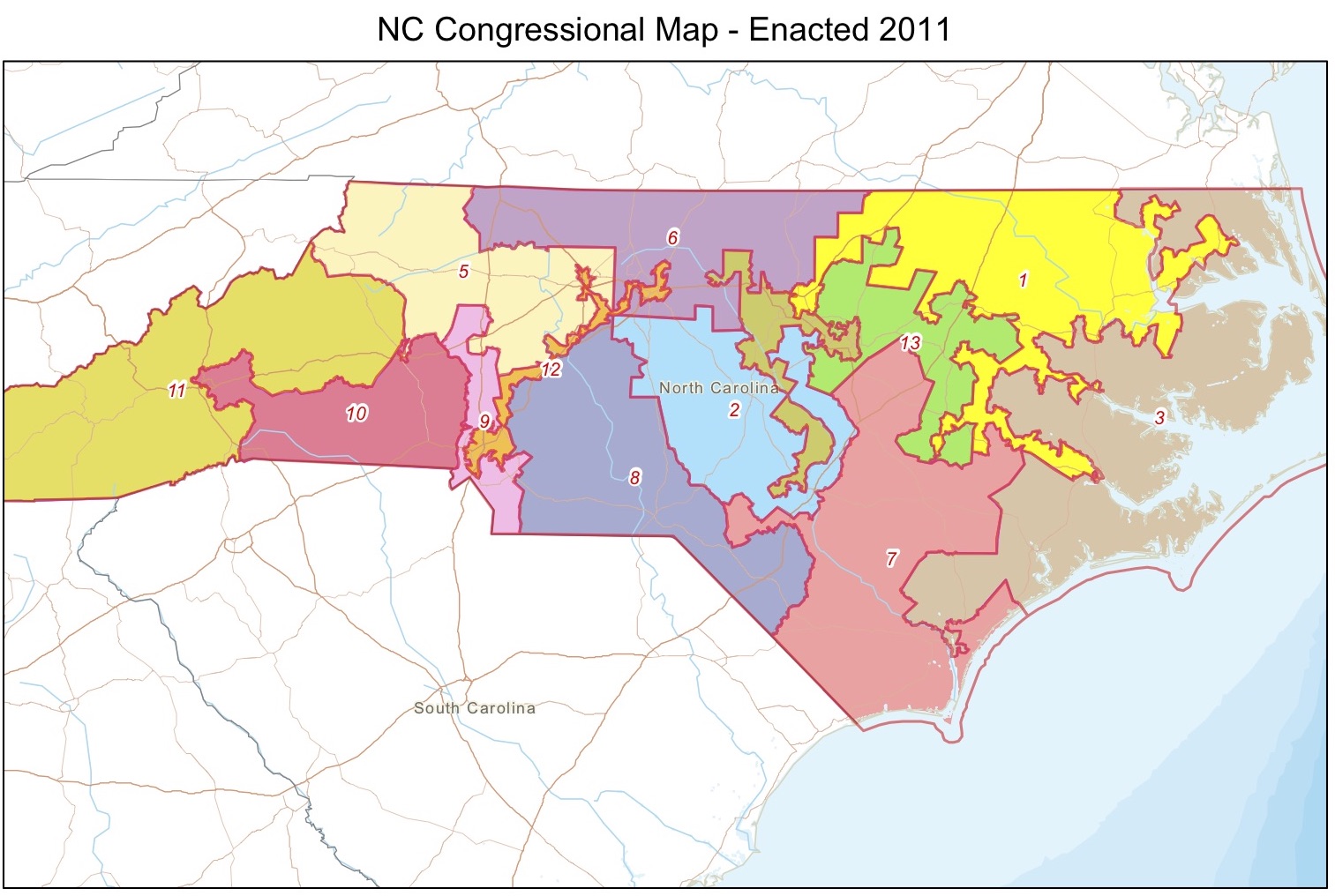}
\caption{2011 NC Congressional districting map}
\label{fig:NCCongressional2011}
\end{figure}

The 2011 North Carolina Congressional districting map \cite{NCDistricts2011} can be seen in Figure \ref{fig:NCCongressional2011}.

The GEO scores for both parties in North Carolina, using the 2011 election districting map and the 2016 Presidential election data, can be seen in Table \ref{table:GEO_NC}.

\begin{table}[h]
\centering
\begin{tabular}{|C{2 cm}||C{1 cm}|C{2.5 cm}|C{2.5 cm}|C{2.5 cm}|} \hline
NC 2016 Presidential & \textbf{$\GEO$} Score & Newly Competitive Districts & Contributing  Stable Wins & Contributing Losses  \\ \hline \hline
Democratic Party & \textbf{6}        &  6, 13, 2, 3, 8, 9  &     12, 1, 4 & 9, 10   \\ \hline
Republican Party  & \textbf{0}           &  (none)        & (none)     & (none)  \\ \hline
\end{tabular}
\caption{GEO scores using North Carolina  2011 districting map and the 2016 Presidential election data.}  
\label{table:GEO_NC}
\end{table}

We note that the districts that are labeled as newly competitive, and contributing districts align with the analysis done by the Quantifying Gerrymandering Group's blog post, ``Towards a Localized Analysis'' \cite{DukeLocalizedAnalysis}.

\begin{figure}[h]
\centering
\includegraphics[width=3 in]{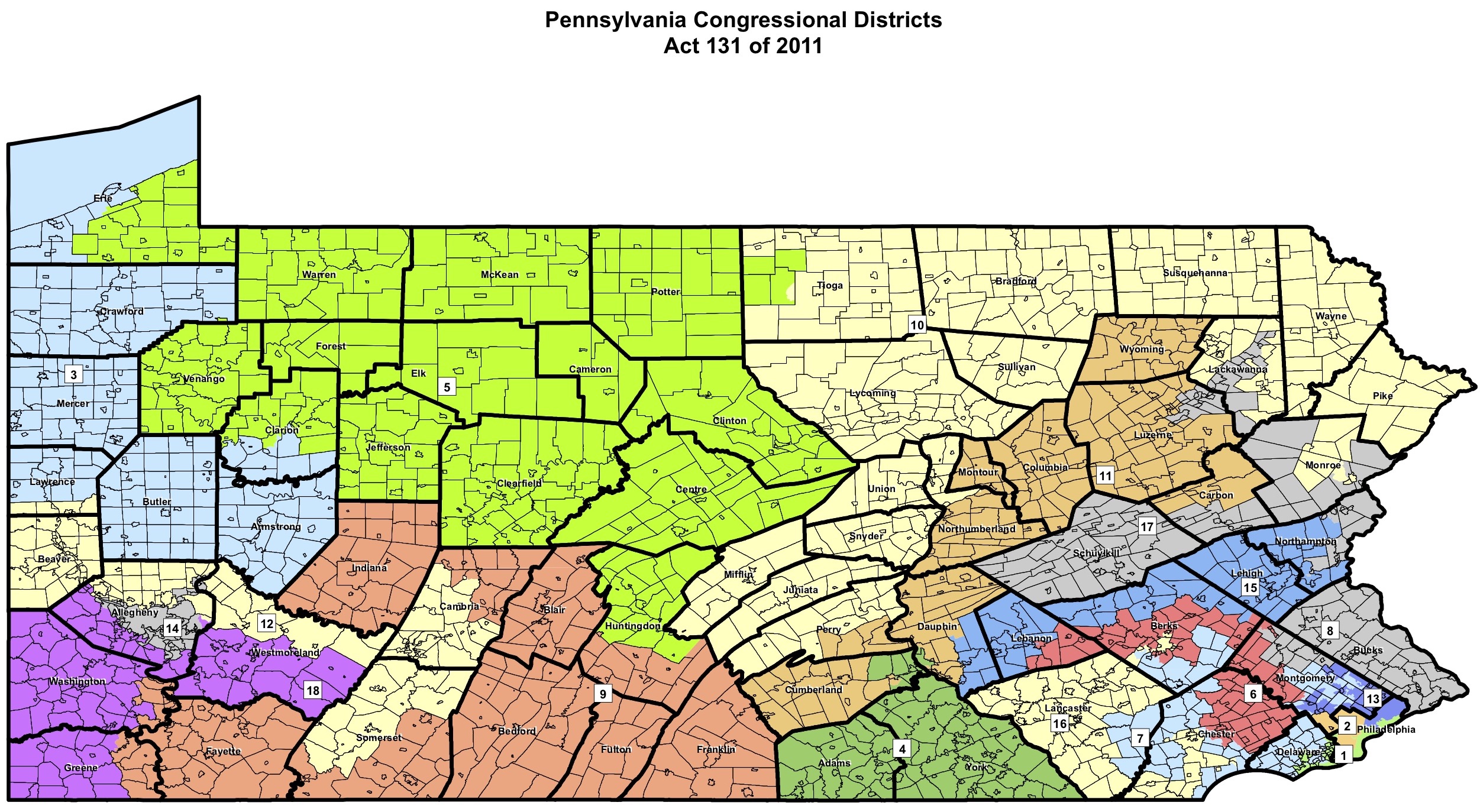}
\caption{2011 PA Congressional districting map}
\label{fig:PACongressional2011}
\end{figure}

The 2011 Pennsylvania Congressional districting map \cite{PADistricts} can be seen in Figure \ref{fig:PACongressional2011}.

The GEO scores for both parties in Pennsylvania, using the 2011 election districting map and the Senate 2016 election outcome data, can be seen in Table \ref{table:GEO_PA}.

\begin{table}[h]
\centering
\begin{tabular}{|C{2 cm}||C{1 cm}|C{2.5 cm}|C{2.5 cm}|C{2.5 cm}|} \hline
Pennsylvania 2016 Senate & \textbf{$\GEO$} Score & Newly Competitive Districts &Contributing  Stable Wins & Contributing Losses   \\ \hline \hline
Democratic Party & \textbf{9}        &  7, 8, 18, 6, 15, 12, 17, 4, 9   &  14, 1, 2, 13 & 5, 3, 11, 16, 10, 9, 17, 15, 12, 4, 6, 8      \\ \hline
Republican Party  & \textbf{0}           &  (none)        & (none)    & (none)   \\ \hline
\end{tabular}
\caption{GEO scores  using Pennsylvania 2011 districting map and the Senate 2016 election outcome data. }
\label{table:GEO_PA}
\end{table}

We note that  the districts that are labeled as newly competitive and contributing districts capture the districts flagged in the analysis done by Azavea in their article, ``Exploring Pennsylvania's Gerrymandered Congressional Districts'' \cite{PAsGerrymanderedDistricts}.  Specifically, that article described districts 1, 13, as Democratically - packed and the GEO flags them as winning contributing districts. They also identify districts 3, 4, 6, 7, 11, 12, 15, 16, 17 as cracking Democratic constituencies and the GEO metric flags them as lost, newly competitive or contributing districts (the contributing districts among those contributing higher vote shares).

\begin{figure}[h]
\centering
\includegraphics[width=3 in]{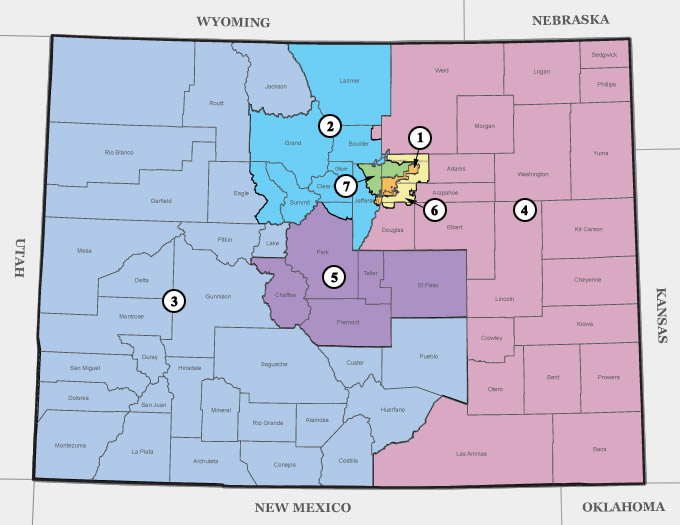}
\caption{2013-enacted CO Congressional districting map}
\label{fig:COCongressional2013}
\end{figure}

The 2013-enacted Colorado Congressional districting map \cite{CODistricts2011} can be seen in Figure \ref{fig:COCongressional2013}.

The GEO scores for both parties in Colorado, using the 2013-enacted districting map and the Governor 2018 election outcome data, can be seen in Table \ref{table:GEO_CO}.  

\begin{table}[h]
\centering
\begin{tabular}{|C{2 cm}||C{1 cm}|C{2.5 cm}|C{2.5 cm}|C{2.5 cm}|} \hline
Colorado 2018 Gubernatorial & \textbf{$\GEO$} Score & Newly Competitive Districts &Contributing  Stable Wins & Contributing Losses   \\ \hline \hline
Democratic Party & \textbf{2}        & 4, 3  &     2, 1, 6, & 3    \\ \hline
Republican Party  & \textbf{1}           &  2        &    4, 5 & 7, 6   \\ \hline
\end{tabular}
\caption{GEO scores  using Colorado's 2013-enacted districting map and the Governor 2018 election outcome data.}  
\label{table:GEO_CO}
\end{table}

Recall that Clelland et al, in their analysis of Colorado \cite{ColoradoInContext}, stated that they ``do not find evidence of effective partisan manipulation in the 2011/2012 adopted maps.''  Nevertheless, they do point out several districts that seemed unusual.  Specifically, in section 5.1 of \cite{ColoradoInContext}, Districts 2, 4, 5, and 7 were singled out for various unusual characteristics. The GEO scores similarly do not show evidence of partisan manipulation as the parties have similar GEO scores. We find it notable, however, that particularly Districts 2 and 4 are singled out by the GEO metric as ``Newly Competitive'' for the Republicans and Democrats respectively, and districts 4, 5 and 7 appear in the Republican party's ``Contributing Districts.''

\section{Maps from the 2021 Redistricting Cycle}
\label{section:newdata}

    Maps that have been drafted following the 2020 Census are newly available, and these are likely the maps of most interest to those who are currently studying redistricting and the detection of gerrymandering.  The Princeton Gerrymandering Project has made a huge number of maps and partisan data available to the public \cite{PGP}, nearly all of which we have evaluated using the GEO metric.  We direct the reader to \censor{ \url{https://www.the-geometric.com/}} for the most up-to-date table of GEO metric analysis of newly released maps, which is growing as additional data are made available. 
    
In Tables \ref{table:2021data1}   and  \ref{table:2021data2}, we give examples of some of the results of our analyses which highlight where the GEO metric disagrees with either the Efficiency Gap, the Declination, or the Mean-Median Difference.  We have circled the metric values that suggest an incorrect conclusion.  For reference, positive values of the Efficiency Gap and Mean-Median indicate a gerrymander benefitting the Democratic party, and positive values of the Declination indicate a gerrymander benefitting the Republican party.  A threshold of $|EG| \geq 0.08$ was suggested for the Efficiency Gap \cite{PartisanGerrymanderingEfficiencyGap}, so we use that threshold here.   No specific threshold value has been given for when  the Declination or Mean-Median Difference indicate a gerrymander, so we considered values which were atypical among all maps analyzed to indicate gerrymanders.  For example, among the nearly 200 maps\footnote{The maps analyzed include several drafts that have been released for some states, which is why the number can be nearly 200.} we analyzed,  about  3.5\% have a Mean-Median Difference above 0.07 in absolute value, and about 15\% had a Declination value above 0.25 in absolute value.

For all the maps in Table \ref{table:2021data1}, the GEO metric indicates that there is no significant partisan bias. The Efficiency Gap suggests that the Massachusetts map is a gerrymander for the Democratic party.  The Mean-Median value for the Maryland map indicates a gerrymander for the Democratic party, while the Efficiency Gap suggests the Maryland map is a gerrymander for the Republican party (again, we direct the reader to \censor{\url{https://www.the-geometric.com/} for additional maps)}.   We believe that the GEO metric gives a more accurate assessment of these maps.  It is worth noting that the work done by the Metric Geometry and Gerrymandering Group \cite{MA_DemDomination} indicates that based on previous data for Massachusetts, ``Though there are more ways of building a valid districting plan than there are particles in the galaxy, every single one of them would produce a 9–0 Democratic delegation.''  Thus it is unlikely that the Massachusetts map referenced in Table \ref{table:2021data1}, which has 8 Democratic seats and 1 Republican seat according to the corresponding partisan data from \cite{PGP}, is a gerrymander for the Democratic party as the Efficiency Gap suggests.  The Princeton Gerrymandering Project gave an ``A'' to the Maryland map \cite{PGP}.

\begin{table}[h]
\centering
\hskip-1.5cm
\begin{tabular}{|M{5 cm}|M{1.7 cm}|M{1.6 cm}|M{1.6 cm}|M{1.6 cm}|M{1.6 cm}|M{1.6 cm}|} \hline
Map & Number of Districts  & Dem GEO & Rep GEO & Declination & Efficiency Gap & Mean-Median \\ \hline\hline
Massachusetts 2021 Draft Staff Congressional Map  &  9  &1 &2 &-0.1077 & \Circle{0.2656}	& -0.0343  \\ \hline
Maryland 2021 Citizens Redistricting Commission Final Draft State Senate Map  &47   & 11 & 10 & 0.0928& \Circle{-0.1044} & \Circle{0.0736}  \\ \hline
\end{tabular}
\caption{Maps where the GEO metric correctly suggests no partisan gerrymandering, in contrast with other metrics.  Metric scores on maps created after the 2021 census.  All data from  \cite{PGP}.  }
\label{table:2021data1}
\end{table}

For the maps in Table \ref{table:2021data2}, the GEO metric does indicate partisan bias (favoring Republicans for Texas and Democrats for both of the Illinois maps).  The Declination suggests no gerrymandering for the Texas map (as do both of the other metrics). The Efficiency Gap suggests no gerrymandering for the Illinois State Senate map (as does the Mean-Median), and the Mean-Median suggests no gerrymandering for the Illinois Congressional map.  Note that the \emph{sign} of the Mean-Median is even incorrect for both of the Illinois maps, as a negative value suggests that those maps are better for the Republican party.   We believe that the GEO metric also gives a more accurate assessment of these maps. The GEO analysis agrees with the analysis done by the Princeton Gerrymandering project \cite{PGP}  where all of these maps receive C to F ratings on partisan fairness with the gerrymander favoring the party indicated by the GEO metric. In addition, many media sources have reported on gerrymandering in those states for this redistricting cycle (see, for example, \cite{TheHillTexas} and \cite{TheHillIllinois}).  

\begin{table}[h]
\centering
\hskip-1.5cm
\begin{tabular}{|M{5 cm}|M{1.7 cm}|M{1.6 cm}|M{1.6 cm}|M{1.6 cm}|M{1.6 cm}|M{1.6 cm}|} \hline
Map & Number of Districts  & Dem GEO & Rep GEO & Declination & Efficiency Gap & Mean-Median \\ \hline\hline
Texas 2021 State House Final Map H2316    & 150 & 46&29 & \Circle{0.0862} &\Circle{-0.0220} & \Circle{-0.0446} \\ \hline
Illinois 2021 Final State Senate Map   &59 & 9 & 18& -0.2582 &\Circle{0.0639} &\Circle{-0.0162} \\ \hline
Illinois 2021 Final Congressional Map    &17 & 1 & 7&-0.4467 &0.1342 & \Circle{-0.0248}  \\ \hline
\end{tabular}
\caption{Maps where the GEO metric correctly suggests partisan gerrymandering, in contrast with other metrics.  Metric scores on maps created after the 2021 census.  All data from  \cite{PGP}.  }
\label{table:2021data2}
\end{table}

It is worth noting that values for the Efficiency Gap, Declination, and Mean-Median tend to be lower for all of the State House and State Senate data we've evaluated and posted on \censor{\url{https://www.the-geometric.com/}.}  This is perhaps not surprising, as those districting maps have higher numbers of districts, so that the sheer numbers of districts can obscure any partisan bias for those particular metrics.  Again, given that the GEO metric represents a \emph{count} (which is straightforward to interpret) we see this as further validation of the utility and accuracy of the GEO metric.

Finally, in Table \ref{table:2021data3}, we give an example of three maps where all metrics suggest that partisan gerrymandering is at play (favoring the Republican party for all three\footnote{As noted above, the Mean-Median Difference \emph{values} may not look large, but they are among the very largest for all of the nearly 200 maps we have analyzed so far.}).  The Princeton Gerrymandering Project analysis \cite{PGP} as well as the media \cite{TheHillTexas, 538NC, WPR_WI} agree that those three maps were gerrymanders.

\begin{table}[h]
\centering
\hskip-1.5cm
\begin{tabular}{|M{5 cm}|M{1.7 cm}|M{1.6 cm}|M{1.6 cm}|M{1.6 cm}|M{1.6 cm}|M{1.6 cm}|} \hline
Map & Number of Districts  & Dem GEO & Rep GEO & Declination & Efficiency Gap & Mean-Median \\ \hline\hline
Texas 2021 Final Congressional Plan C2193 & 38 & 13 & 5 & 0.2517 & -0.0910 & -0.0879 \\ \hline
North Carolina 2021 CST-13 Final Congressional Map \newline (HB 977/SB 740) & 14 & 5 & 1 & 0.4022 & -0.1992 & -0.0631 \\ \hline
Wisconsin 2021 State Legislative Congressional Draft Plan SB622 & 8 & 4 & 0 & 0.5757 & -0.2649 & -0.0687 \\ \hline
\end{tabular}
\caption{Maps where the all metrics agree on partisan gerrymandering.  Metric scores on maps created after the 2021 census.  All data from  \cite{PGP}.  }
\label{table:2021data3}
\end{table}

\section{GEO Metric Analysis}
\label{section:analysis}

Many analyses of metrics intended to detect partisan gerrymandering have centered on instances in which the metric is equal to 0, as this is the ``ideal'' value of the metric \cite{2018arXiv180105301V}, \cite{DeclinationAsMetric}, \cite{ImplementingPartisanSymmetry}.   We do not consider a GEO metric score of 0 to be more desirable than nonzero GEO scores that are relatively balanced in each party.  Indeed $GEO_P = 0$ indicates that party $P$ has \emph{no reasonable room for improvement}, suggesting that the map is designed to benefit party $P$.\footnote{It is worth noting here that we do not consider the value of $GEO_P - GEO_Q$ to be as useful as knowing both $GEO_P$ and $GEO_Q$.  Certainly reporting both values gives more information that is lost by simply reporting $GEO_P-GEO_Q$, one can easily compute $GEO_P-GEO_Q$ if both of those values are calculated, and knowing that one party's GEO metric score is close to or equal to 0 suggests that the outcome could not reasonably be improved for that party.} So we focus our analysis on what properties would contribute to a larger GEO score for party $P$.

Using the notation from Section \ref{section:defns}, let's suppose that district $D_1$ contributes to party $P$'s GEO score.  Say that the neighboring districts of $D_1$ that party $P$ lost are $D_2, D_3, \dots, D_k$, and the neighboring districts that party $P$ won are $D_{k+1}, D_{k+2}, \dots, D_m$.  Furthermore, suppose that  $D_{k+1}, D_{k+2}, \dots, D_\ell$ are the districts whose vote share is only allowed to go down to $A_i - \sigma$.  That is, 
\begin{equation*}
A_i - \sigma > 0.55
\end{equation*}
While  $D_{\ell+1}, D_{\ell+2}, \dots, D_{m}$ are the districts whose vote share is only allowed to go down to $0.55$.  That is, 
\begin{equation*}
A_i - \sigma \leq 0.55
\end{equation*}

Then, since $D_1$ contributes to party $P$'s GEO score, we must have that, if $V_i^*$ is the current recorded vote share for district $D_i$ when district $D_1$ is considered in the algorithm,\footnote{Note that the moment when a district is encountered in the algorithm impacts whether or not it contributes to the GEO metric.}
\begin{align}
0.5 - V^*_1 &< \sum_{\substack{i=2 \\ V^*_i > A_i - \sigma}}^\ell \left(V^*_i- \left(A_i - \sigma \right)\right) + \sum_{\substack{j=\ell+1 \\ V^*_j > 0.55}}^m \left(V^*_j-0.55\right)     \nonumber \\
&= N\sigma +  \sum_{\substack{i=2 \\ V^*_i > A_i - \sigma}}^\ell \left(V^*_i-A_i  \right) + \sum_{\substack{j=\ell+1 \\ V^*_j > 0.55}}^m \left(V^*_j-0.55\right)
\label{eqn:GEOcontribution}
\end{align}
where $N$ is the size of the set $\{i: 2 \leq i \leq \ell,  V^*_i > A_i - \sigma\}$.   Certainly, the left hand side of Equation \eqref{eqn:GEOcontribution} is small (making the equation more likely to be true) if $V^*_1$ is close to 0.5.  So what makes the right hand side of Equation \eqref{eqn:GEOcontribution} large?  

Certainly, if there are many packed districts for party $P$, then the sum  $\sum_{\substack{i=k+1 \\ V^*_i > A_i-\sigma}}^\ell (V^*_i-A_i) +  \sum_{\substack{j=\ell+1 \\ V^*_j > 0.55}}^m \left(V^*_j-0.55\right)$ will be large.  If there are many districts whose vote share is somewhat large, compared to the average neighborhood vote share, then the sum $\sum_{\substack{i=2 \\ V^*_i > A_i - \sigma}}^k\left(V^*_i-A_i  \right) $ will be larger.  This arguably gets at where party $P$ is cracked.  

How about the number $N \sigma$?  If district $D_1$ has more neighbors, then $N$ could potentially be larger (as well as the other sums in Equation \eqref{eqn:GEOcontribution}).  And $\sigma$ is larger if the standard deviation of the $A_i$ is large.  

We summarize this discussion in terms of vote shares.  District $D_1$ is more likely to contribute to party $P$'s GEO score if:
\begin{enumerate}
\item Party $P$ is packed in nearby districts
\item Party $P$ is cracked in nearby districts
\item  District $D_1$ has many neighbors
\item  There is large variation in the neighborhood average vote shares $A_1, A_2, \dots, A_n$
\end{enumerate}

Items (1) and (2) are of course desired, but it's worth discussing whether (3) and (4) are intuitive and/or desired.  A district having many neighbors certainly could indicate an irregularly drawn district.  For example, Pennsylvania's 7th district from the map in Figure \ref{fig:PACongressional2011} (the so-called `Goofy Kicking Donald Duck District'') has many neighbors, arguably because of the way that the districts have been cut around it in order to increase the number of Republican-won districts.  Having many neighbors may indeed indicate something unusual in the district drawing.  But a thorough analysis of the relationship between gerrymandering and districts incident with many neighbors has not, to our knowledge, been explored.  The relationships between districts with many neighbors, gerrymandering, and the GEO metric are worth further exploration.  

Having a large variation among the vote shares $A_1, A_2, \dots, A_n$ could certainly mean that districts are intentionally drawn to be far from the mean (by packing, for example).  It could also simply be a result of having two sections of the state which are both geographically separated, and also politically polarized.  Or it could be the result of very politically polarized regions in the state, and the districts are drawn along the lines of partisan polarization.  In this way, one could argue that the GEO score is likely to be higher in a politically polarized state.  The precise ways in which this plays out are also worth exploring. No metric is meant to be a stand alone measure of gerrymandering. Where the GEO indicates potential gerrymandering we recommend further analysis. 

We finally note that if $\sigma = 0$, then $A_i = \frac{\sum_{j=1}^n A_j}{n}$ for each $i$.  In other words, the state is extremely homogeneous politically, which makes it very difficult for a mapmaker to draw a map to benefit any party.  In that situation, it is expected that the GEO metric would be quite low.

\section{Using GEO metric with ensembles}
\label{section:MCMC}

In the past five years or so, mathematicians have promoted the usage of \emph{outlier analysis} for the purpose of detecting gerrymandering.  See, for example, \cite{RamachandranGoldOutlierAnalysis} for an overview of the outlier analysis method.  We briefly describe this method as follows:  a large number of potential districting maps is created; the set of such maps is called an \emph{ensemble}.  All maps in the ensemble satisfy that state's set of restrictions, whether they include Voting Rights Act requirements, compactness requirements, or any other state-specific requirements.  A proposed map is then compared to all other maps in the ensemble.  This comparison can be made using any kind of metric.  For example, we could use a single set of election data and simply see how many districts the Republican, or Democratic, party would have won with each map in the ensemble (in this example the metric is simply number of seats won).  The proposed map can be compared with all maps in the ensemble by seeing how unusual \emph{the proposed map's} number of Republican seats is.  That is, we can see if the proposed map's number of Republican seats is typical, unusually high, or unusually low as compared with the number of Republican seats in all maps in the ensemble.

There are a variety of ensemble creation methods that have been promoted; because of the mathematical theory and rigor behind them, we focus on ensemble creation methods that use a Markov Chain Monte Carlo (MCMC) process.  For examples of the kinds of MCMC algorithms that have been proposed for the purpose of creating an ensemble of districting maps, see \cite{DukeWisconsin},  \cite{DukeNC}, \cite{RecomMGGG}, \cite{DukeMergeSplit}, \cite{DukeMultiScaleMergeSplit}.

While the GEO metric does take both geographic and election outcome data into account, it does not  look at the actual locations of voters to see if the vote share swaps incorporated in calculating the GEO metric are physically possible.  The creation of an ensemble of maps \emph{does} create a wide variety of allowable maps, and thus enhances the utility of the GEO metric by allowing us to compare a map's GEO metric to the GEO metric of many other allowable maps.  We used the Metric Geometry Gerrymandering Group's publicly available GerryChain Recom MCMC \cite{RecomMGGG} to create an ensemble of maps for each of  North Carolina, Pennsylvania and Colorado's 2011 maps.    We followed the description parameters set up at \cite{GerryDetails}.    We also took 10,000 steps in the chain for each map.  

\begin{figure}[h]
	\centering
	\subcaptionbox{ Democratic GEO score \label{fig:NCDem}}
         {\includegraphics[width=2.2 in]{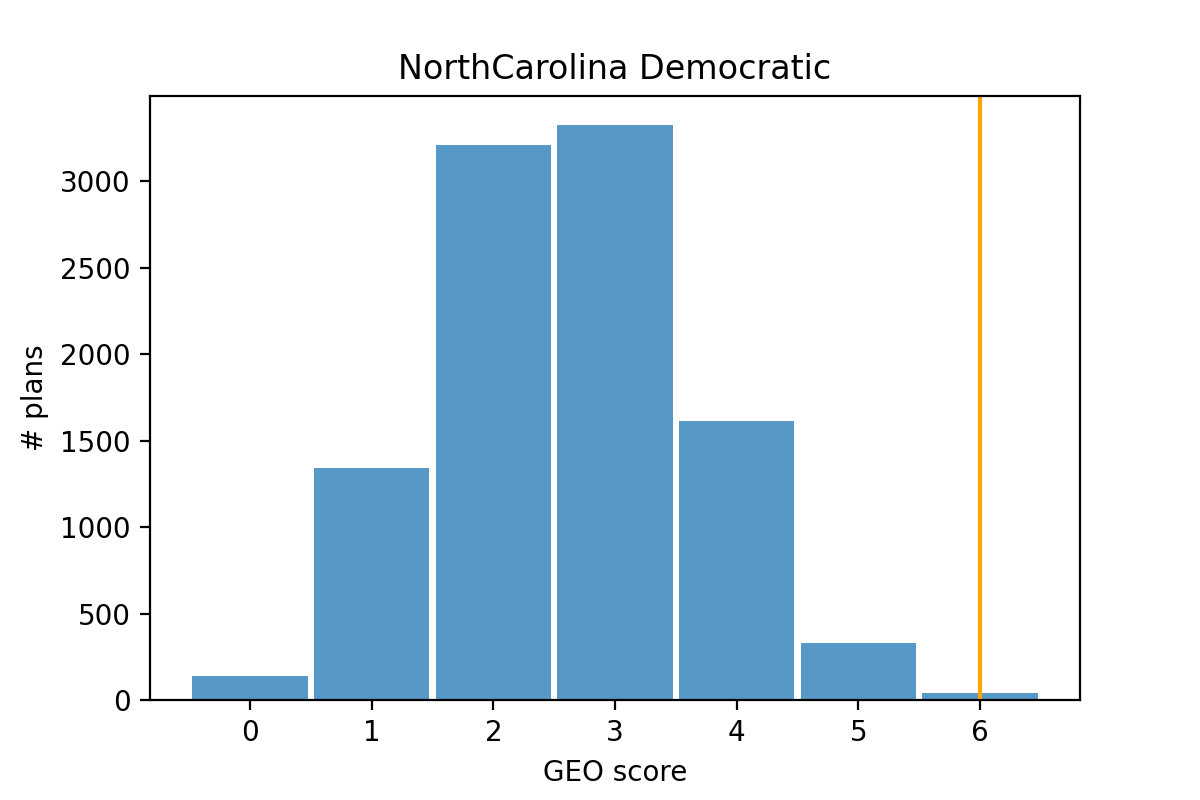}}
       \subcaptionbox{Republican GEO score \label{fig:NCDem}}
         {\includegraphics[width=2.2 in]{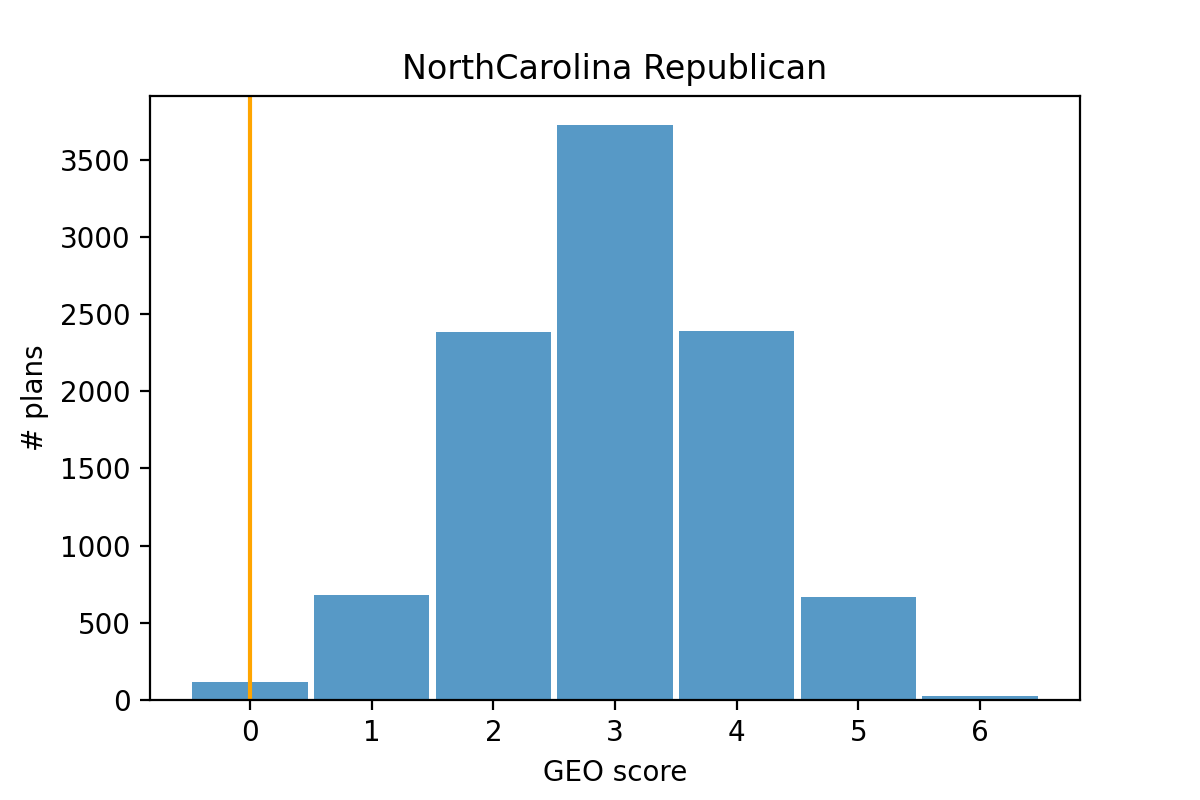}}
        \caption{Ensemble outcomes for North Carolina, using the 2016 Presidential election outcome data.  The yellow line is the corresponding value for the 2011 Congressional redistricting map.}
        \label{fig:EnsembleNC}
\end{figure}

\begin{figure}[h]
	\centering
		\subcaptionbox{ Democratic GEO score \label{fig:PADem}}
         {\includegraphics[width=2.2 in]{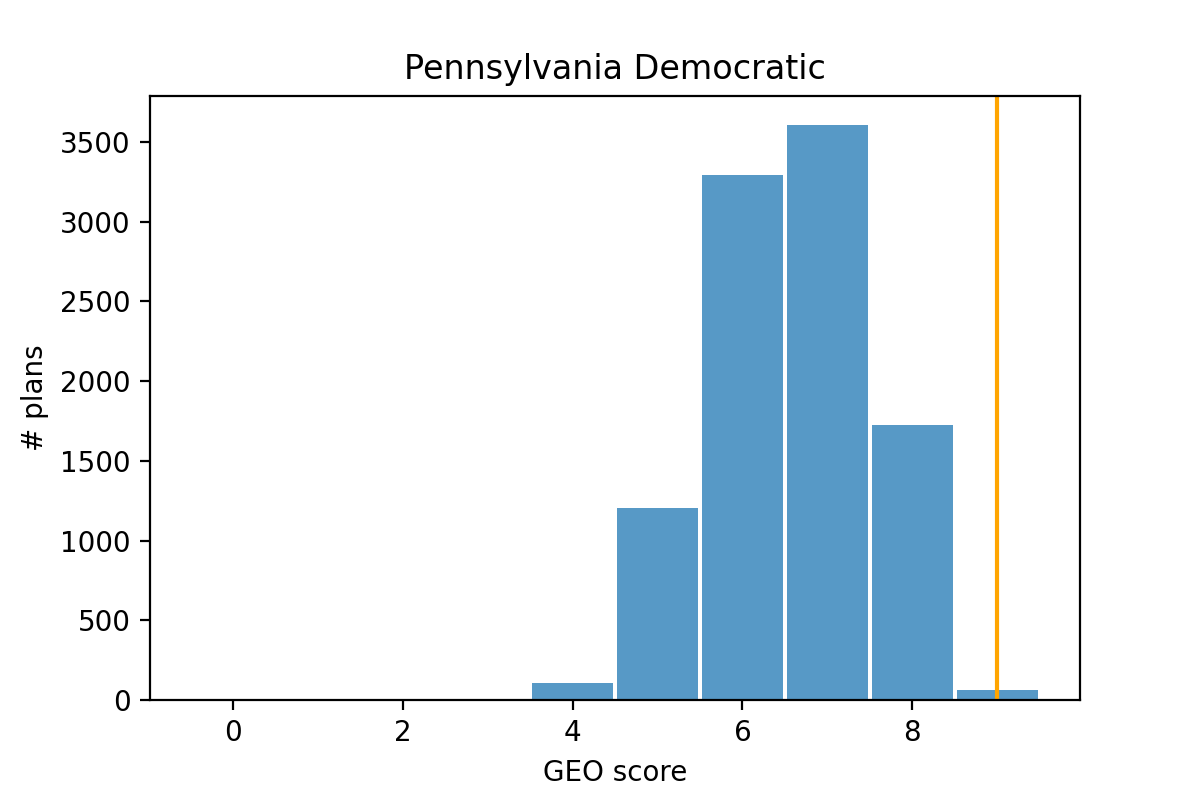}}
	\subcaptionbox{ Republican GEO score \label{fig:PADem}}
         {\includegraphics[width=2.2 in]{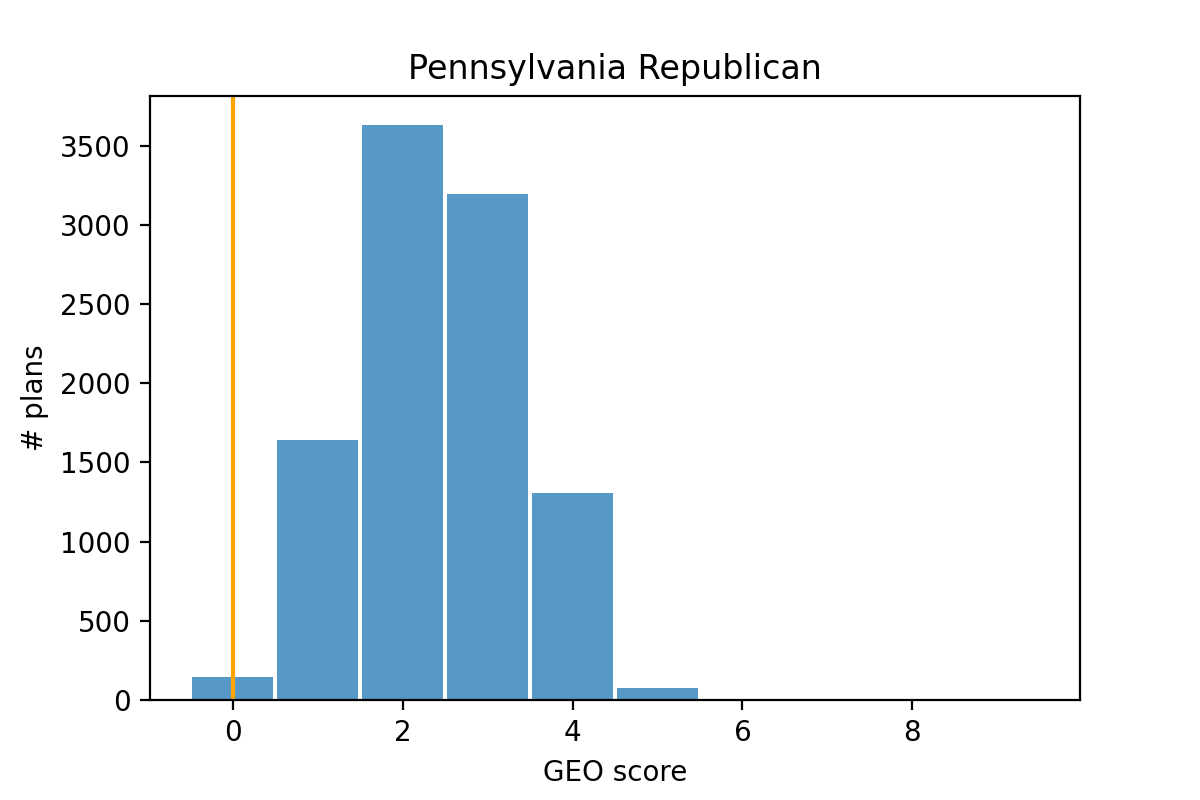}}
        \caption{Ensemble outcomes for Pennsylvania, using the Senate 2016 election outcome data.  The yellow line is the corresponding value for the 2011 Congressional redistricting map.}
        \label{fig:EnsemblePA}
\end{figure}

\begin{figure}[h]
	\centering
	\subcaptionbox{ Democratic GEO score \label{fig:CODem}}
         {\includegraphics[width=2.2 in]{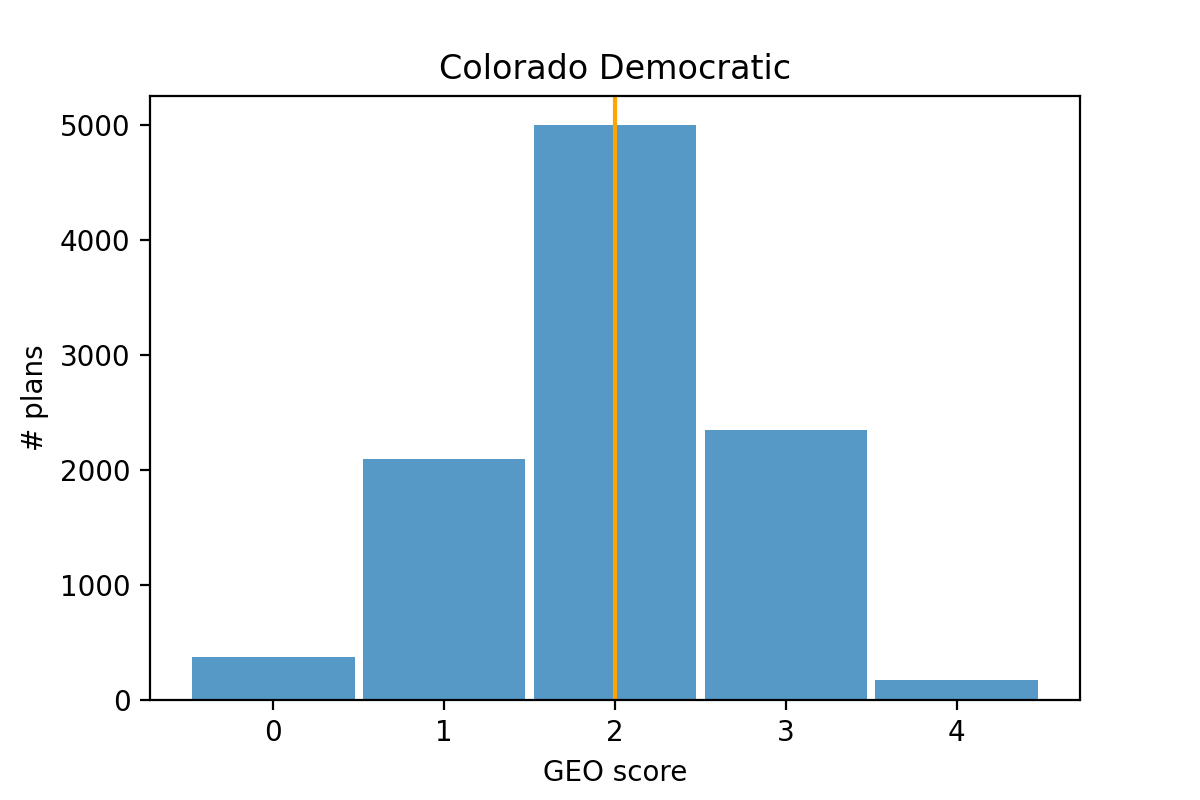}}
       \subcaptionbox{Republican GEO score \label{fig:CODem}}
         {\includegraphics[width=2.2 in]{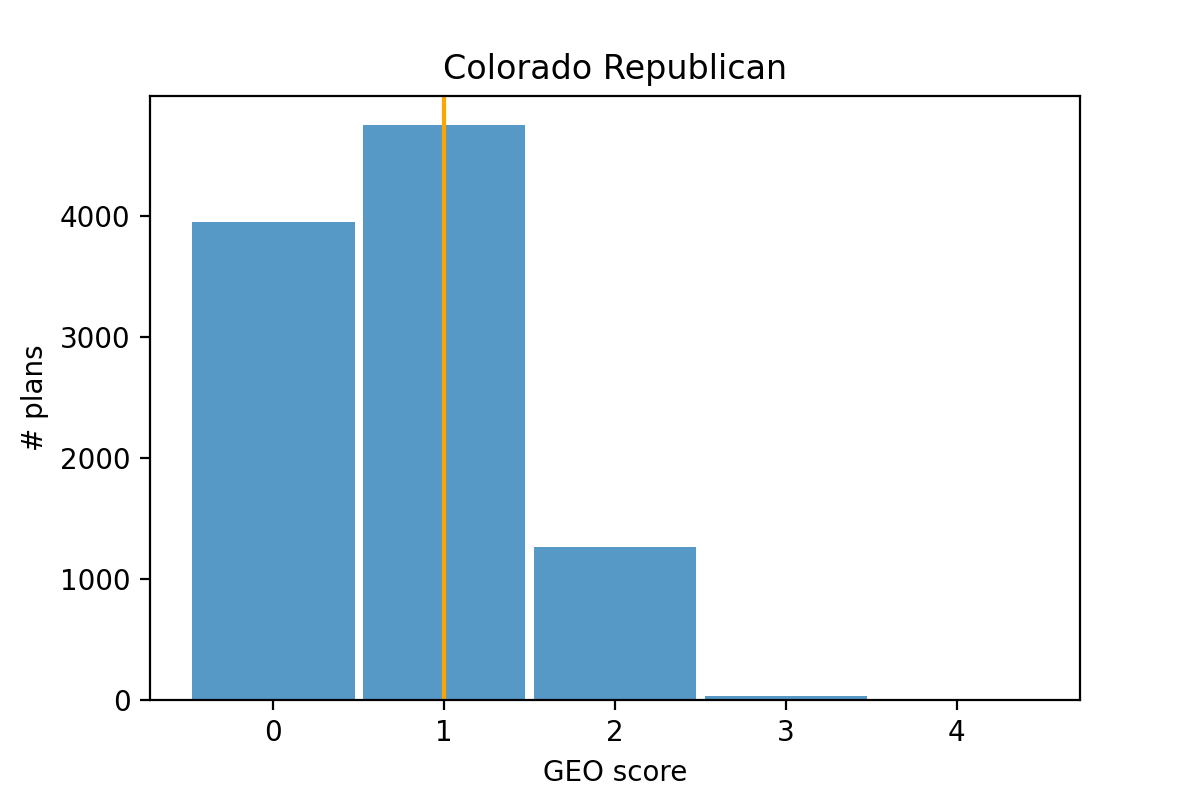}}
        \caption{Ensemble outcomes  for Colorado, using the Governor 2018 election outcome data.  The yellow line is the corresponding value for the 2013-enacted Congressional redistricting map.}
        \label{fig:EnsembleCO}
\end{figure}

The outcome of this outlier analysis can be seen in Figures \ref{fig:EnsembleNC}, \ref{fig:EnsemblePA}, and \ref{fig:EnsembleCO}.  

As expected, we can see that the Democratic GEO metric scores in both North Carolina and 
Pennsylvania are unusually high, while the Republican GEO metric scores in those states 
are unusually low.  And the GEO score for each party in Colorado is fairly typical within their ensembles, as expected.

\section{Caveats, Clarifications, and Takeaways} \label{section:CCT}

The big idea behind the GEO metric is to detect when boundary lines between two districts could potentially be adjusted so that a political party might gain an additional seat without risking any of its current seats.  This is achieved by considering which districts are adjacent, and looking at the vote shares of those adjacent districts.  The metric does \emph{not} look at the actual locations of voters to see if the vote share swaps incorporated in calculating the GEO metric are physically possible, and thus does \emph{not} propose a specific alternative map.  So while it can suggest that a better outcome for a particular party seems likely, it cannot guarantee that such a better outcome is available.

\begin{figure}[h]
	\centering
	\subcaptionbox{ Party $P$ districts won \label{fig:StateXDistrictsWon}}
         {\includegraphics[width=2.2 in]{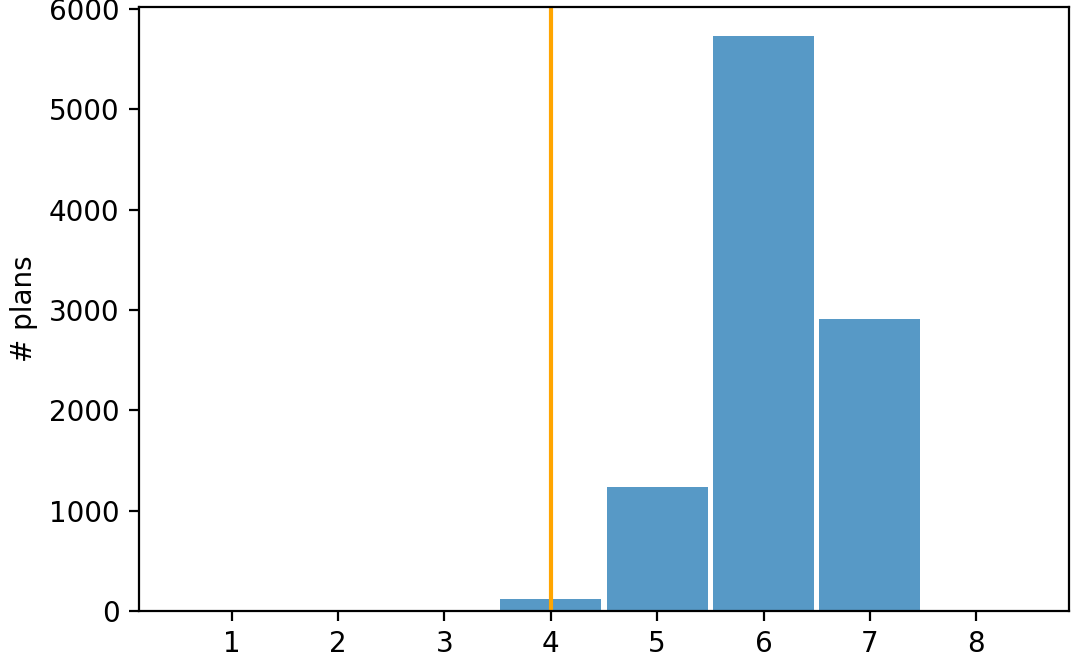}}
       \subcaptionbox{Efficiency Gap \label{fig:StateXEG}}
         {\includegraphics[width=2.2 in]{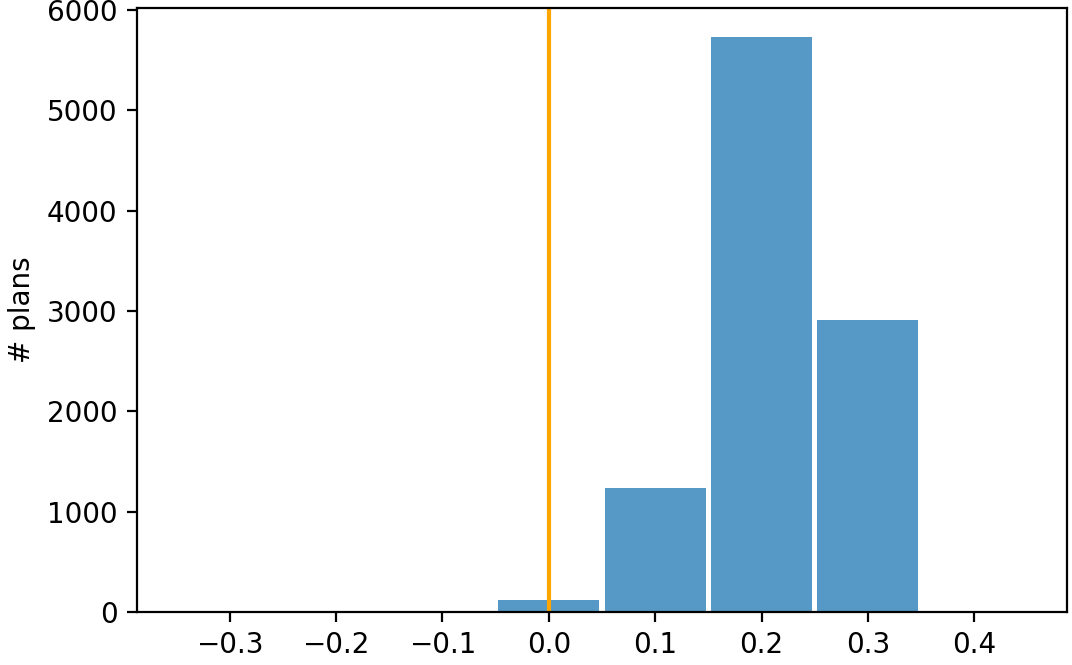}}
         \subcaptionbox{Mean-Median Difference \label{fig:StateXMeanMedian}}
         {\includegraphics[width=2.2 in]{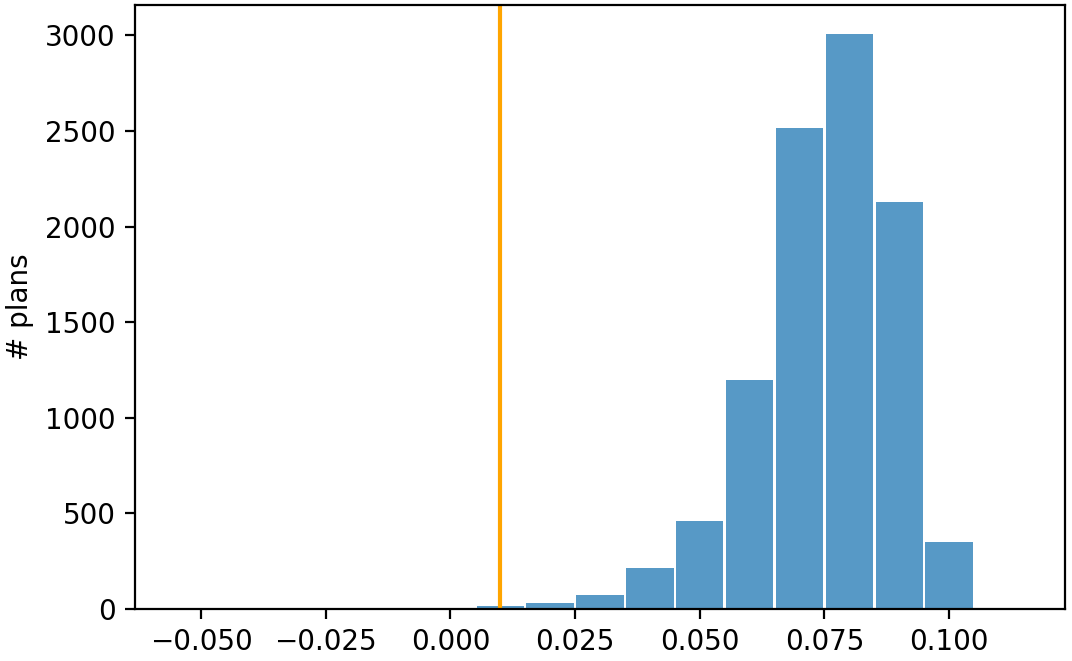}}
         \subcaptionbox{GEO Metric, Party $P$ \label{fig:StateXGEOA}}
         {\includegraphics[width=2.2 in]{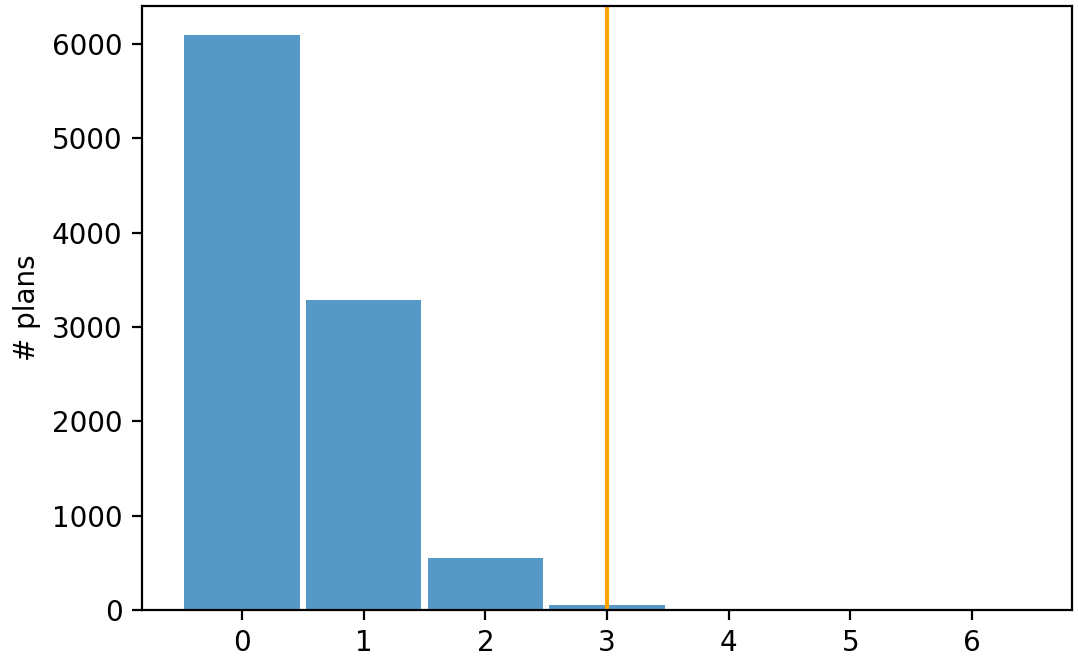}}
          \subcaptionbox{GEO Metric, Party $Q$ \label{fig:StateXGEOB}}
         {\includegraphics[width=2.2 in]{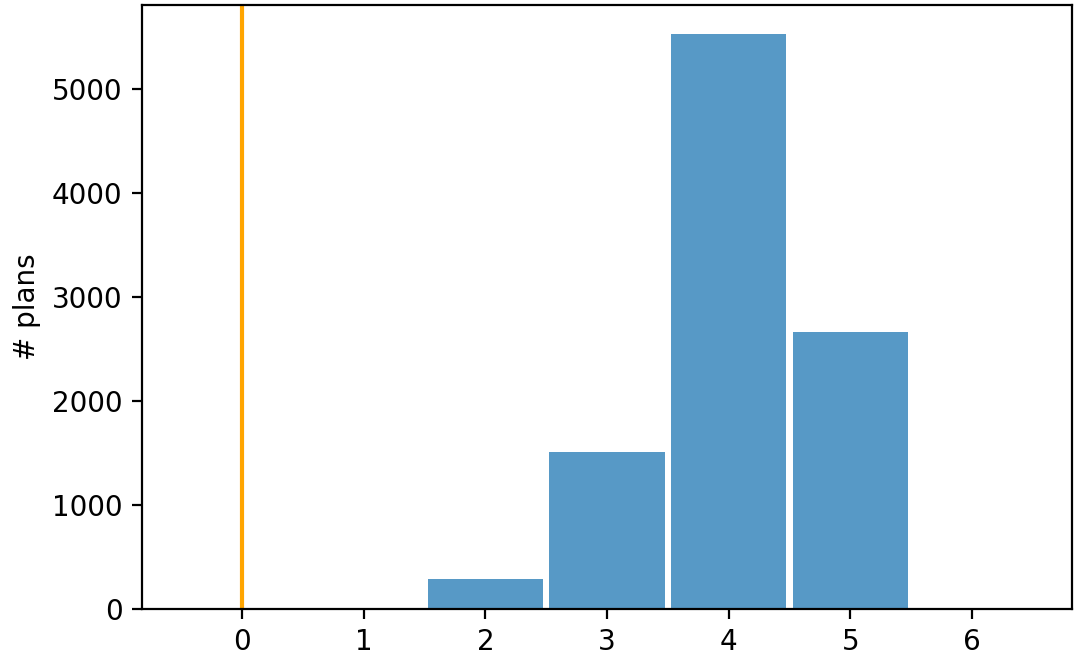}}
        \caption{Ensemble outcomes  for State $X$.}
        \label{fig:EnsembleStateX}
\end{figure}

\begin{figure}[h]
	\centering
	\subcaptionbox{ Party $P$ districts won \label{fig:StateYDistrictsWon}}
         {\includegraphics[width=2.2 in]{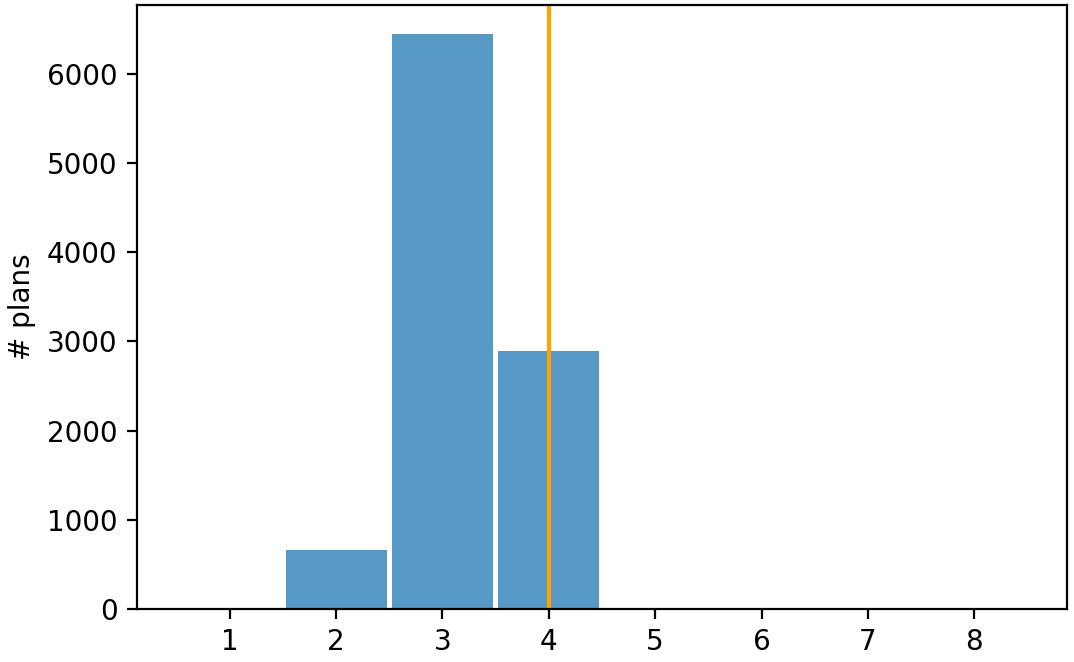}}
       \subcaptionbox{Efficiency Gap \label{fig:StateYEG}}
         {\includegraphics[width=2.2 in]{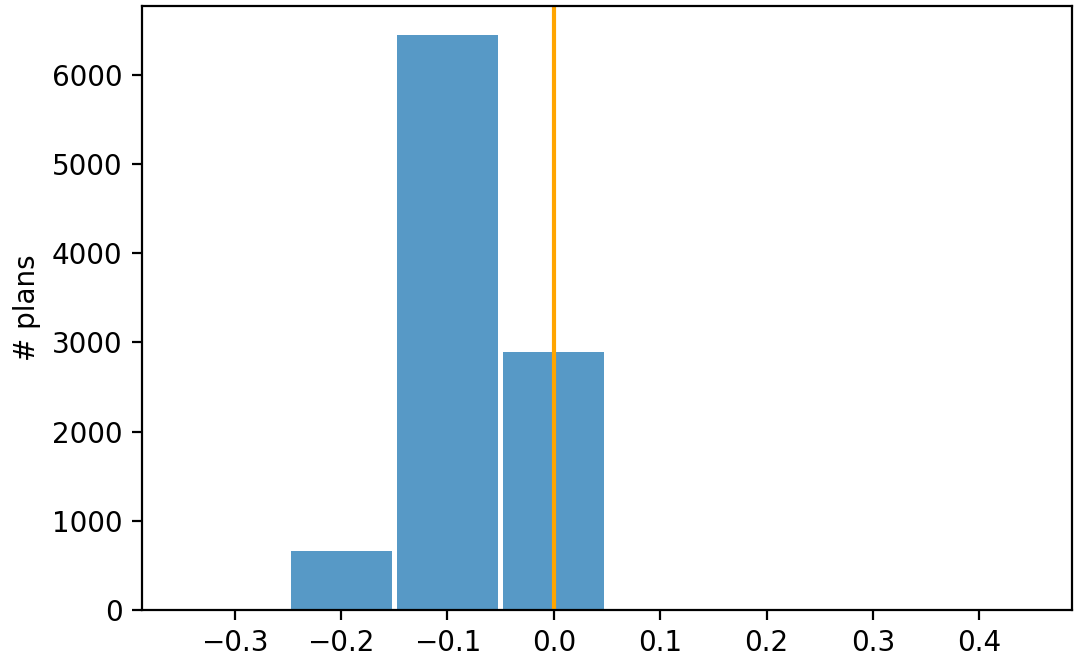}}
         \subcaptionbox{Mean-Median Difference \label{fig:StateYMeanMedian}}
         {\includegraphics[width=2.2 in]{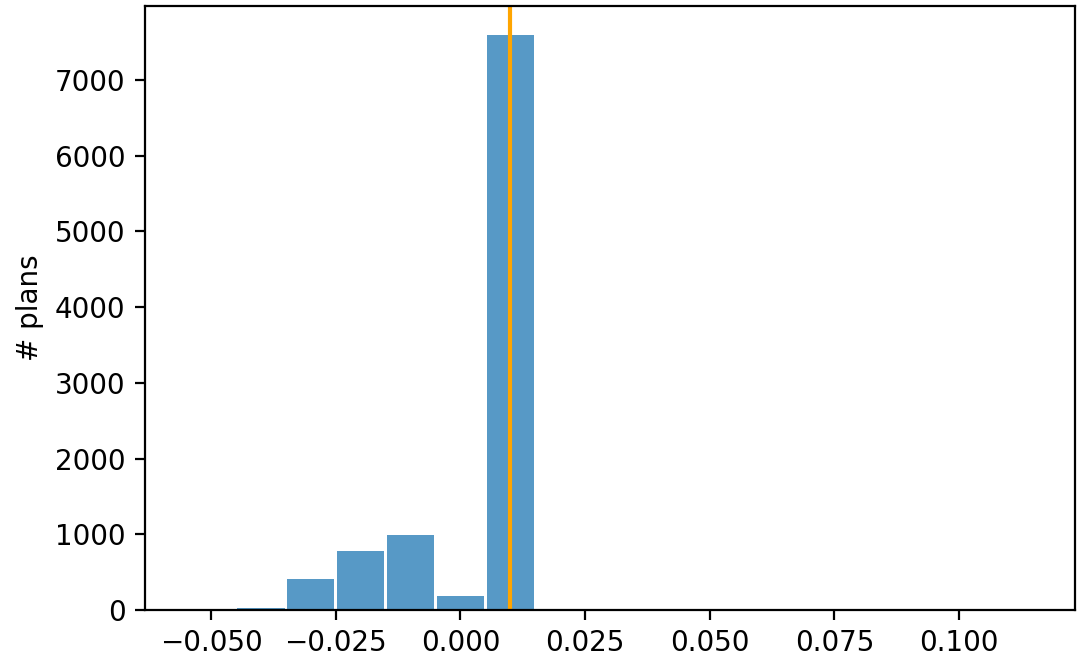}}
         \subcaptionbox{GEO Metric, Party $P$ \label{fig:StateYGEOA}}
         {\includegraphics[width=2.2 in]{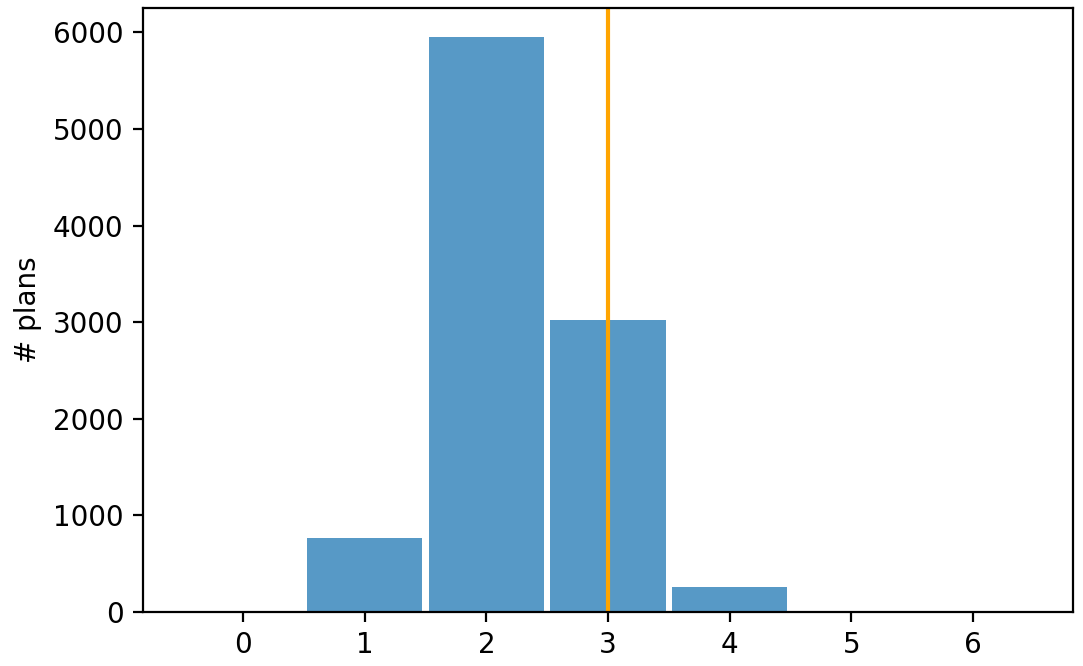}}
          \subcaptionbox{GEO Metric, Party $Q$ \label{fig:StateYGEOB}}
         {\includegraphics[width=2.2 in]{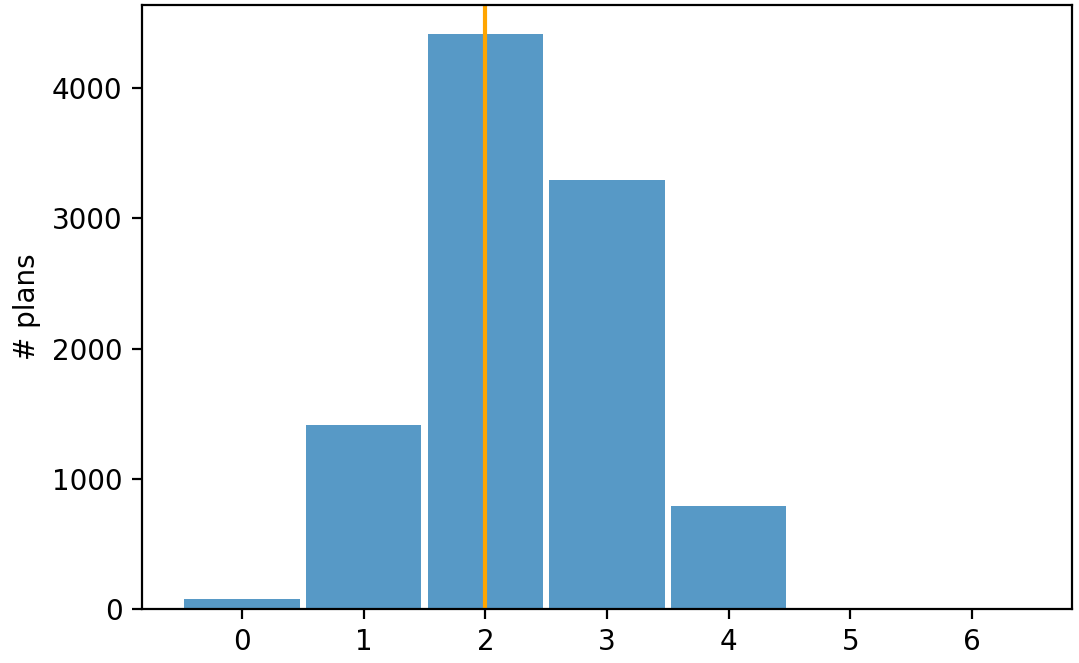}}
        \caption{Ensemble outcomes  for State $Y$.}
        \label{fig:EnsembleStateY}
\end{figure}

The ensemble method produces achievable maps which can be compared to a proposed map, indicating how typical or atypical a proposed map is, given that state's political landscape.  However the choices enacted in the ensemble sampling strategy impact which maps are sampled (potentially introducing bias in the sample) and result in a nondeterministic outcome.  We believe that the GEO metric can achieve much of what the ensemble method can achieve, but without any potential sampling bias.  Furthermore, we believe that the \emph{value} of the GEO metric is much more useful than other highly-utilized metrics, like the Mean-Median Difference and the Efficiency Gap.  This was discussed in Section \ref{section:newdata}, but we can also show this for our sample states states $X$ and $Y$ from Section \ref{subsec:MotivatingExample}.  In Figures \ref{fig:EnsembleStateX} and \ref{fig:EnsembleStateY} we can see that the ensemble distributions for State $X$ for all metrics indicate that this state is potentially gerrymandered.  And the ensemble distributions for State $Y$ for all metrics suggest that State $Y$ is likely not gerrymandered.  However, the Mean-Median Difference and Efficiency Gap \emph{values} for both states are identical, and suggest no gerrymandering in both states.  It's only the GEO metric \emph{values} for State $X$ that indicates potential gerrymandering.

The significance of a particular GEO metric value is highly dependent on the number of districts in a state.  Thus, when evaluating the GEO metric values for different parties within a state, one should also consider the number of districts.  A GEO metric score of 5 for party $P$ and 0 for party $Q$ is much more concerning in a state with 10 districts than in a state with 100 districts.  We've chosen to keep the GEO metric score as a \emph{count} (by not dividing by the number of districts, for example) because we'd like the value to have more meaning than simply ``this map appears to be gerrymandered.''  Specifically,the GEO metric score is an indication of how many more districts could have potentially been won by a party.

However, we emphasize that the goal of the GEO metric is not to declare the number of additional districts that a party \emph{should have} won, but rather the number of additional districts a party \emph{could have} won.  In general, because the algorithm behind the GEO metric changes the district's vote share to be 50/50, it is indeed most  appropriate to say that party $P$ could have won about $\frac{\GEO_P}{2}$ additional
districts; the idea being that after the transferring of vote shares, party $P$ has a 50/50 chance of winning each of the ``newly competitive'' districts.  We chose \emph{not} to have the algorithm behind the GEO metric swap vote shares in order to give party $P$ a safe win because we didn't want to advocate for a party $P$ gerrymander.  Rather, we'd like to see how much potential party $P$ has for improvement.  We note that, for all of the outlier analyses we did, each party $P$ did have maps that achieve $\left\lfloor \frac{\GEO_P}{2} \right\rfloor$ additional districts for that party.\footnote{Within the respective ensembles, 23\% of CO's maps, 36\% of NC's maps, and 1\% of PA's maps achieved an additional  $\left\lfloor \frac{\GEO_\text{Dem}}{2} \right\rfloor$ districts for the Democratic party; the low percentage for PA is likely due to a higher $\sigma$ in that state (please note the future research questions outlined in this section).  And 77\% of CO's maps, 2\% of NC's maps, and 2\% of PA's maps achieved an additional $\left\lfloor \frac{\GEO_\text{Rep}}{2} \right\rfloor$ districts for the Republican party; the low percentages for NC and PA are likely because $\GEO_{Rep} = 0$ in those states, suggesting the map is already ``optimized'' for the Republican party.  We direct the reader to \censor{\url{https://www.the-geometric.com/}}, as our GEO metric calculations there indicate that $\left\lfloor \frac{\GEO_P}{2}\right\rfloor$ additional seats is achievable, based on the ensemble analyses that The Princeton Gerrymandering Project completed for many of those maps.}

This idea of potential for improvement of each party is the best way to think about the GEO metric.  If, for example, a state has 15 districts, and we know that $\GEO_P = 5$, while $\GEO_Q = 0$, this indicates that party $P$ could potentially have a much better outcome, while party $Q$ has no flexibility to improve its outcome.  This lack of flexibility for party $Q$ indicates that the map may have been drawn to optimize party $Q$'s outcome.  Whereas, if a state has 15 districts and we have $\GEO_P =5$, and $\GEO_Q = 4$, both parties have flexibility to improve their outcome.  Because it focuses on this presence of flexibility, the GEO metric does a better job than other metrics of determining when a party is potentially the beneficiary of gerrymandering.  Specifically, if a party’s GEO score is 0, this indicates a lack of flexibility in the map to improve that party’s outcome.

The GEO metric was designed to utilize the state-specific nuances in partisan makeup and map data.  These nuances also invite additional research, including the following topics: 
\begin{enumerate}
\item Unstable win parameter $w$: In this paper we set the parameter $w$=0.55. This defines the category ``unstable win'' as districts with vote shares between 0.5 and 0.55. Other ranges may be more suitable for certain states. Further research is needed to determine which factors should be used to determine the unstable win range for a particular state.
\item The relationship between districts with many neighbors and gerrymandering: as stated in Section \ref{section:analysis}, a district having many neighbors could increase its ability to contribute to the GEO score.  It is of interest to know how the number of neighbors that districts have can impact the GEO score.
\item Standard deviation of the average neighbor vote share:  in Section \ref{section:analysis} we noted that a large $\sigma$ could contribute to a higher GEO score.  What more can we say about the relationship between $\sigma$ and the GEO scores for a state?   Is it possible to differentiate when high $\sigma$ values are due to natural partisan makeup of the state versus gerrymandering?
\end{enumerate}

In summary, the GEO metric is an improvement on prior metrics.  It uses both the Geography of the map and Election Outcome data to detect the presence of gerrymandering.   The GEO metric is a fixed deterministic calculation that does not rely on sampling method choices or hidden probability distributions, and thus has the potential for wider acceptance in the courts.  As we have seen, the \emph{value} of the GEO metric has more meaning than the values of metrics like the Efficiency Gap or Mean-Median Difference.  The GEO metric is a count of the number of districts that could have become competitive for each party, under reasonable changes to the map.  Whereas the Efficiency Gap and Mean-Median difference values have no meaning unless compared with other maps in an ensemble.  There are no fixed threshold values that we promote in order to determine exactly when gerrymandering has happened, but a reasonable comparison of the GEO metric score for each party, taking into account the total number of districts,  will indicate the potential for improvement in that party's outcome with the given election outcome data. 

\section*{Acknowledgements}
The authors would also like to thank the referees for their careful reading and suggestions, which have made the paper much stronger.

\censor{
The authors would also like to thank Ben Harte for contributing to software development in this project, and the Fletcher Jones funding that supported him.  Dr. Veomett was supported by funding from the Saint Mary's College of California Provost's Research Grant and Dr. Campisi was supported by funding from the San Jose State University Central RSCA Grant.
}

\bibliographystyle{plain}   
\bibliography{GEOmetric}

\end{document}